\documentclass[%
10pt,tightenlines,
superscriptaddress,
onecolumn,
nofootinbib,
 amsmath,amssymb,
 aps,
pra,
]{revtex4-2}

\usepackage[margin=0.8in]{geometry}
\usepackage[T1]{fontenc}

\usepackage{graphicx}
\usepackage{dcolumn}
\usepackage{bm}
\usepackage{color,xcolor}
\usepackage[colorlinks,citecolor=blue]{hyperref}
\usepackage{CJKutf8}
\usepackage{braket}
\usepackage{amsthm}
\usepackage{amsmath}
\usepackage{physics}
\usepackage[most]{tcolorbox}
\usepackage{titletoc}
\usetikzlibrary{arrows.meta,decorations.pathreplacing}
\definecolor{lblue}{HTML}{CCE7F1}
\definecolor{green}{HTML}{D9EDCE}
\definecolor{purple}{HTML}{E3D7EF}
\definecolor{gold}{HTML}{C99228}
\definecolor{pink}{HTML}{C54778}

\newtheorem{theorem}{Theorem}[section]
\newtheorem{lemma}[theorem]{Lemma}
\newtheorem{definition}[theorem]{Definition}
\usepackage{bbm}
\usepackage{makecell}
\usepackage{enumitem}
\usepackage{cancel}
\usepackage{amsthm}

\newtcolorbox[auto counter]{mybox}[2][]{
	enhanced,
	breakable,
	colback=blue!5!white,
	colframe=blue!75!black,
	fonttitle=\bfseries,
	title=Box \thetcbcounter: #2,#1
}

\begin{document}
\begin{CJK}{UTF8}{gbsn}

\title{Learning Many-Body Hamiltonians Using a Local Probe}

\author{Suying Liu} 
\thanks{Equal contribution (listed alphabetically).}
\affiliation{Joint Center for Quantum Information and Computer Science, University of Maryland, College Park, Maryland 20742, USA}
\affiliation{Department of Computer Science, University of Maryland, College Park, USA}

\author{Zitai Xu} 
\thanks{Equal contribution (listed alphabetically).}
\affiliation{Joint Center for Quantum Information and Computer Science, University of Maryland, College Park, Maryland 20742, USA}
\affiliation{Department of Computer Science, University of Maryland, College Park, USA}

\author{Alexey V. Gorshkov} 
\affiliation{Joint Center for Quantum Information and Computer Science, University of Maryland, College Park, Maryland 20742, USA}
\affiliation{Joint Quantum Institute, NIST/University of Maryland, College Park, MD, 20742, USA}

\author{Xiaodi Wu} 
\affiliation{Joint Center for Quantum Information and Computer Science, University of Maryland, College Park, Maryland 20742, USA}
\affiliation{Department of Computer Science, University of Maryland, College Park, USA}

\author{Yu-Xin Wang (王语馨)}
\email{yxwang.physics@outlook.com}
\affiliation{Joint Center for Quantum Information and Computer Science, University of Maryland, College Park, Maryland 20742, USA}

\author{Zhi-Yuan Wei (魏志远)} 
\email{zywei@umd.edu}
\affiliation{Joint Center for Quantum Information and Computer Science, University of Maryland, College Park, Maryland 20742, USA}
\affiliation{Joint Quantum Institute, NIST/University of Maryland, College Park, MD, 20742, USA}

\date{\today}

\begin{abstract}
Hamiltonian learning provides a systematic framework for reconstructing unknown quantum dynamics. However, existing protocols typically assume direct measurement access to the entire system. With fast single-qubit control and a connected reference backbone, we show that a single measurable qubit suffices to learn all $O(N)$ independent parameters of a bounded-degree two-body Hamiltonian on $N$ qubits at the Heisenberg limit. Crucially, our protocol uses robust SWAP gates synthesized by quantum signal processing, enabling coherent transfer of states evolving under distant Hamiltonian parameters to the measurable qubit. This transfer requires no prior calibration of the Hamiltonian parameters of the intermediate links. A parallel learning architecture achieves total query time $\widetilde{O}(N)$ for an $N$-qubit chain, while retaining Heisenberg-limited precision scaling. On the chain, these scalings match the fundamental precision and information-propagation lower bounds up to logarithmic factors. The framework further extends to arbitrary bounded-degree interaction graphs. Our results establish a scalable route to learning an extensive number of many-body Hamiltonian parameters through only a local measurement interface.
\end{abstract}

\maketitle
\end{CJK}

\section{Introduction}
\label{sec:introduction}
Complex dynamical systems are rarely observed in their entirety. Across the classical sciences, the properties of an extended system are routinely inferred from a limited set of spatially local observations. Weather forecasting combines measurements from a network of monitoring stations to reconstruct and predict atmospheric dynamics; seismology infers subsurface structure from signals recorded at a finite collection of sensors; and system identification seeks to determine the internal dynamics of an unknown system through a small number of input--output ports. These examples illustrate a broad principle of scientific inference: observations obtained through a limited local interface can nevertheless reveal properties of a much larger system. This principle raises a natural question for quantum many-body systems:
\begin{quote}
    \emph{To what extent can global quantum dynamics be learned through
    a local observation window?}
\end{quote}
We ask this question in the setting of Hamiltonian learning. Suppose that an $N$-qubit system evolves under an unknown local Hamiltonian containing $O(N)$ independently varying parameters, while measurements are available only on a single qubit. Can the Hamiltonian throughout the system still be reconstructed? If so, how efficiently can information about spatially distant Hamiltonian terms be extracted through such a fixed local interface?

Hamiltonian learning provides a systematic framework for reconstructing the generator of unknown quantum dynamics from experimental data. It plays an important role in calibrating and verifying quantum devices, characterizing analog quantum simulators, and probing naturally occurring many-body systems~\cite{BaireyAradLindner2019,HaahKothariTang2024,huang2023learning}. Recent advances have established efficient learning from high-temperature thermal states and real-time evolution~\cite{HaahKothariTang2024}, with subsequent developments extending thermal-state learning to arbitrary fixed temperatures and improving its temperature dependence~\cite{BakshiEtAl2024Thermal,Narayanan2025Flat}. For learning from dynamics, coherent control enables Heisenberg-limited precision scaling~\cite{huang2023learning}, with further progress encompassing unknown interaction structures, compressed-sensing methods, and ansatz-free reconstruction~\cite{BakshiEtAl2024Structure,ma2024learning,hu2025ansatz,LiuWuNiu2025}. Recent protocols also achieve Heisenberg-limited scaling using only static single-qubit control fields~\cite{Brahmachari2026}. These approaches, however, typically assume measurement access distributed throughout the system. Retaining these learning capabilities when measurements are confined to a fixed small subsystem remains a distinct challenge.

Restricted readout is a natural regime in several quantum architectures. A particularly direct example is provided by solid-state spin registers, where one optically or electrically addressable spin serves as the primary readout ancilla for a larger register of nearby nuclear spins, whose information is accessed indirectly by coherently transferring it to the measured degree of freedom \cite{TaminiauEtAl2014,Abobeih2019,vandeStolpe2024}. A similar separation between processing and measurement arises in reconfigurable neutral-atom architectures, where coherent single-qubit operations can be performed across large arrays but global measurements are comparatively slow. Recent developments coupling neutral atoms to optical cavities can accelerate measurements within small, dedicated readout regions \cite{Bluvstein2024,Deist2022}. Similar local readout-zone architectures also appear in trapped-ion systems \cite{PinoEtAl2021}. In these latter platforms, physical shuttling can itself transport atoms or ions to the probing regions and thus provides an additional routing primitive. More broadly, these examples motivate treating the spatial extent of measurement access as an experimental resource distinct from the size of the quantum system itself.

Motivated by this setting, we study an asymmetric access model with fast single-qubit control throughout the system and evolution under its unknown native Hamiltonian, while direct state preparation, measurement, and reset are confined to one distinguished \emph{bright qubit}. We assume that each neighboring pair has a specified interaction component, which we call a reference interaction, with an unknown strength bounded away from zero, ensuring that information can propagate along the entire chain to the measurement site. Under this model, we show that a single measurable qubit suffices to reconstruct all local fields and two-body couplings of a general nearest-neighbor Hamiltonian on an $N$-qubit chain, achieving the Heisenberg-limited $O(1/\varepsilon)$ evolution-time scaling for target precision $\varepsilon$ and total evolution time $\widetilde{O}(N)$ at fixed precision and confidence.

Figure~\ref{fig:overview} gives an overview of our protocol. Our approach combines parallel phase encoding with robust information transfer through unknown interactions. Hamiltonian reshaping first isolates disjoint local terms, allowing many coefficients to be encoded simultaneously into separate qubit phases. These records are then conveyed to the measurable endpoint and read out sequentially, avoiding the quadratic overhead of probing each distant interaction in a separate experiment. Crucially, the transfer operations must themselves be constructed from the unknown Hamiltonian. Using quantum signal processing, we synthesize robust SWAP operations from the reference interactions, without prior learning of individual coupling strengths or signs. We bound the accumulated transfer and measurement errors so that robust phase estimation retains Heisenberg-limited precision scaling. The approach extends to Hamiltonians having bounded-degree interaction graphs with a connected reference backbone. On chains with a common-axis reference interaction, the protocol requires only a restricted form of global control, removing the need for individual addressing of the unmeasured qubits. 

The remainder of the paper is organized as follows. Section~\ref{sec:related-work} compares our results with previous local-probe experiments and restricted-access Hamiltonian-learning protocols. Section~\ref{sec:setup} defines the learning task, the control and readout model, and the reference-backbone assumptions, and reviews Hamiltonian reshaping and robust phase estimation. Section~\ref{subsec:benchmark-scalings} presents the fundamental precision and information-propagation lower bounds for the learning task. Section~\ref{sec:robust} develops the parallel learning protocol, including the robust SWAP construction and parallel learning architecture, and analyzes its error and runtime guarantees. It also presents an implementation with restricted global control. Section~\ref{sec:general-graphs} extends the protocol to bounded-degree interaction graphs. Section~\ref{sec:discussion} concludes with a discussion of interaction geometry, experimental constraints, and open questions. Detailed proofs are provided in the Supplementary Material.

\begin{figure}[htbp]
    \centering
    \input{figure/overview-tikz-v2}
    \caption{\textbf{Overview of the learning protocol.} (a) Time runs upward. Repeatedly resetting the bright qubit and applying robust SWAPs prepares the register approximately in the all-zero state. The unknown Hamiltonian parameters are then encoded in parallel into quantum states distributed across the system. Alternating SWAP layers transport this information back toward the bright qubit, where the encoded states are read out sequentially. (b) The SWAP operations are constructed directly from the unknown interactions using quantum signal processing. Combining robust rotations along three axes produces a SWAP that works uniformly over the promised range of interaction strengths, without prior calibration.}
    \label{fig:overview}
\end{figure}

\section{Related work}
\label{sec:related-work}
Previous studies have also used a local probe to reveal interactions within an extended quantum system. Abobeih et al.~\cite{Abobeih2019} used a single nitrogen--vacancy centre to characterize a cluster of $27$ nuclear spins, and van de Stolpe et al.~\cite{vandeStolpe2024} mapped a $50$-spin network through correlated sensing. These experiments reconstruct interactions in specific solid-state spin registers. Our result addresses the corresponding algorithmic question for general local fields and two-body Pauli couplings. We give an explicit total-evolution-time guarantee, including initialization and information transport through unknown interactions, with near-linear system-size dependence on chains and Heisenberg-limited precision scaling.

Restricted-access Hamiltonian identification has also been established for structured spin models. Burgarth, Maruyama, and Nori~\cite{BurgarthMaruyamaNori2009} reconstructed coupling strengths in spin chains from access to an end spin, while Burgarth and Maruyama~\cite{BurgarthMaruyama2009Gateway} extended identification to networks with known topology and a suitable accessible gateway. Di Franco, Paternostro, and Kim~\cite{DiFrancoPaternostroKim2009} reconstructed spin-chain couplings from single-spin measurements without initializing the rest of the chain. Sone and Cappellaro~\cite{SoneCappellaro2017} analyzed single-probe identifiability for specified interaction models and estimated the required sampling and evolution resources. These results establish reconstruction under particular interaction and access assumptions, but do not provide the combined near-linear system-size and Heisenberg-limited precision guarantee for all coefficients of a general two-body chain Hamiltonian obtained here. Our protocol learns spatially varying local fields and all two-body Pauli components, and explicitly accounts for the cost of conveying their signals to a single fixed readout qubit.

More recently, Chen, Cotler, and Huang~\cite{chen2025quantumprobetomography} introduced quantum probe tomography and proved efficient reconstruction of translation-invariant nearest-neighbor lattice Hamiltonians from single-site probes of thermal states, up to inversion symmetry about the probe. Their algorithm has query complexity polynomial in $1/\varepsilon$. Our protocol requires neither translational invariance nor thermal-state preparation, and reconstructs an extensive set of independent coefficients. It achieves Heisenberg-limited precision scaling and, at fixed target precision, learns the chain using a local probe in a total time of $\widetilde O(N)$, matching the information-propagation lower bound up to logarithmic factors.

\section{Setup and preliminaries}
\label{sec:setup}

\subsection{System and learning task}
\label{subsec:Hamiltonian-model}
One-dimensional chains and two-dimensional lattices are standard interaction geometries for quantum devices~\cite{Zhang2017,Bernien2017,Philips2022}. In the following discussion, we focus on the 1D chain model, which has the sparsest connectivity and the largest graph diameter among these fixed-size regular geometries. The chain model therefore represents the most stringent setting in terms of the propagation distance required to transfer information from remote interactions to the locally accessible qubit. Our approach extends naturally to higher-dimensional regular lattices, where the increased connectivity provides additional pathways for information propagation and can reduce the learning cost. We discuss the extension to general interaction graphs in Sec.~\ref{sec:general-graphs}.

Label the qubits consecutively by $1,\ldots,N$ and take the first qubit to be the measurable qubit.  We consider the most general nearest-neighbor two-body Hamiltonian,
\begin{equation}
    H(\boldsymbol\lambda)=\sum_{i=1}^{N}\sum_{\alpha\in\{X,Y,Z\}}h_i^\alpha \sigma_i^\alpha+
    \sum_{i=1}^{N-1}\sum_{\alpha,\beta\in\{X,Y,Z\}}J_{i,i+1}^{\alpha\beta}\sigma_i^\alpha \sigma_{i+1}^\beta,
    \label{eq:parallel-general-H}
\end{equation}
where $\sigma_i^\alpha$ denotes the Pauli $\alpha$ operator acting on qubit $i$ and the identity elsewhere. The real coefficients $h_i^\alpha$ describe local fields, and $J_{i,i+1}^{\alpha\beta}$ describe interactions between adjacent qubits. Each qubit has three local-field components, and each of the $N-1$ neighboring pairs has nine interaction components, corresponding to the choices of $\alpha$ and $\beta$. The full parameter vector $\boldsymbol\lambda$ therefore contains $3N+9(N-1)=12N-9$ coefficients. Components known to be absent may be set to zero. An identity term is omitted because it contributes only an unobservable global phase.

Given an accuracy $\varepsilon>0$ and a failure probability
$\delta\in(0,1)$, our goal is to construct an estimator
$\widehat{\boldsymbol\lambda}$ satisfying
\begin{equation}
    \Pr\!\left[
        \|\widehat{\boldsymbol\lambda}
          -\boldsymbol\lambda\|_\infty
        \leq\varepsilon
    \right]
    \geq 1-\delta.
    \label{eq:learning-objective}
\end{equation}
Thus, with probability at least $1-\delta$, every unknown
coefficient is estimated to additive error at most
$\varepsilon$. The available experimental operations and
the prior information about the Hamiltonian are specified
below.

\subsection{Control and readout model}
\label{subsec:access-model}

Coherent control and measurement need not have the same spatial availability. For instance, in solid-state spin-qubit systems, electron spins can be addressed directly through optical initialization and readout, whereas the surrounding nuclear spins are accessible only indirectly, via their hyperfine coupling to the electron. In those systems, efficient control of the nuclear spins typically requires precise knowledge of their interaction Hamiltonian with the electron~\cite{TaminiauEtAl2014,Abobeih2019,vandeStolpe2024}. Related spatial asymmetries arise when neutral-atom or trapped-ion architectures concentrate fast, high-fidelity measurement in dedicated cavity volumes or readout zones~\cite{Deist2022,PinoEtAl2021}. We consider a chain with
coherent single-qubit control at every site, while initial state preparation, measurement and reset are available only at the first qubit. We call this qubit the \emph{bright qubit} and the remaining qubits the \emph{dark register}. Specifically, the available operations are: 

\begin{enumerate}[label=(B\arabic*)]
    \item\label{assumption:reset} preparation and repeatable reset of the bright qubit to the state $\ket{0}$;
    \item\label{assumption:control} arbitrary single-qubit unitaries at every site;
    \item\label{assumption:evolution} evolution under $H(\boldsymbol\lambda)$ for chosen durations, interleaved with the allowed controls; and
    \item\label{assumption:measurement} computational basis measurement of the bright qubit.
\end{enumerate}
Together, \ref{assumption:reset} and \ref{assumption:control} allow preparation of the bright qubit in any pure single-qubit state. The dark register, however, may initially be in an unknown mixed state. \ref{assumption:control} and \ref{assumption:measurement} allow measurement of the bright qubit in any single-qubit basis.  Control choices may depend on previous measurement outcomes, but not on unknown coefficients except through information obtained from those outcomes or specified in advance.

Throughout this work, we assume that the durations of single-qubit gates, bright qubit state preparation, measurement and reset are negligible compared with the total evolution time under the unknown Hamiltonian. Thus, the protocol cost is given by the total Hamiltonian evolution time over all shots. If the evolution intervals have durations $t_1,\ldots,t_M$, this resource is
\begin{equation}
    T_{\rm tot}:=\sum_{m=1}^{M}t_m.
    \label{eq:total-time-resource}
\end{equation}
Evolution of the full register for time $t$ contributes $t$, and repeating the same experiment $R$ times contributes $Rt$.

For simplicity, we first assume independent single-qubit control throughout the chain (assumption (B2) above). This assumption is stronger than necessary. In Sec.~\ref{subsec:global-control-main}, we show that the protocol remains efficient under a substantially more restricted global-control model.

\subsection{Connectivity and prior information}
\label{subsec:transport-assumptions}

We collect here all promises on the unknown Hamiltonian.  First, we assume a known conservative upper bound \(\Lambda\) on the Hamiltonian coefficients:
\begin{equation}
|h_i^\alpha|\leq\Lambda,
\qquad
|J_{i,i+1}^{\alpha\beta}|\leq\Lambda.
\label{eq:coefficient-bound}
\end{equation}
Such a bound can be obtained from the designed operating range of the device or from coarse preliminary calibration and does not require prior knowledge of any individual coefficient.

Second, some interaction must physically connect every dark qubit to the bright qubit. If all interaction coefficients across a bond vanish, the chain would separate into two noninteracting subsystems, and measurements on the first qubit could reveal nothing about the disconnected part. In engineered quantum devices, the physical form of the intended interaction is generally known from the design. Although the exact coefficients on bonds may be unknown, the device parameters and coarse calibration usually provide a range for their strengths. We therefore assume known constants $J_{-},J_{+}>0$ such that, on every bond \((i,i+1)\), one specified Pauli component has coefficient inside this range. More precisely, for known \(a_i,b_i\in\{X,Y,Z\}\), we define
\begin{equation}
    P_i^{\rm ref}=\sigma_i^{a_i}\sigma_{i+1}^{b_i},
    \qquad
    g_i:=J_{i,i+1}^{a_i b_i}.
    \label{eq:transport-component}
\end{equation}
The \emph{reference-backbone assumption} is then
\begin{equation}
    J_{-}\leq|g_i|\leq J_{+},\qquad 1\leq i<N.
    \label{eq:basic-transport-assumption}
\end{equation}
The $g_i$ may vary from bond to bond, and their magnitudes are unknown inside the window bounded by $J_{-},J_{+}$.  The promise says only that the device contains a known chain of bounded interactions. All other coefficients in Eq.~\eqref{eq:parallel-general-H} remain unrestricted within Eq.~\eqref{eq:coefficient-bound}. The lower bound is needed to obtain a uniform runtime guarantee: without it, an intermediate coupling could be arbitrarily weak, creating a bottleneck that makes the time required to access spins and interactions beyond that link arbitrarily long.

\subsection{Hamiltonian reshaping}
\label{subsec:reshaping-main}

A measured signal generally depends on several terms of the Hamiltonian. It is useful to simplify this dependence by averaging away unwanted terms using fast single-qubit pulses, a procedure called \emph{Hamiltonian reshaping}~\cite{huang2023learning}. Its basic mechanism is sign reversal. For example, conjugation by $X$ changes $Z$ to $-Z$, so equal periods of evolution under $hZ$ and its conjugate cancel that term.

More generally, write $H=\sum_a\lambda_aP_a$ and choose a distribution $\mathcal D$ of product Pauli pulses $R$. Interleaving sufficiently short evolution intervals with these pulses realizes an effective Hamiltonian given by the average over $\mathcal D$,
\begin{equation}
    H_{\mathcal D}
    =\mathbb E_{R\sim\mathcal D}[R H R^\dagger]
    =\sum_a\lambda_a\mathbb E_{R\sim\mathcal D}[\chi_R(P_a)]P_a,
    \label{eq:main-reshaped-hamiltonian}
\end{equation}
where $RP_aR^\dagger=\chi_R(P_a)P_a$ and $\chi_R(P_a)\in\{+1,-1\}$. Terms whose signs are equally often positive and negative average to zero. Terms commuting with every pulse retain their original coefficients.

We now describe the reshaping used to isolate several disjoint Pauli terms simultaneously. Let $\{P_a\}_{a=1}^m$ be the terms to be retained, with pairwise disjoint supports $S_a\subseteq[N]$. For simplicity we assume all $P_{a}=\prod_{j\in S_a}Z_j$ are purely $Z$-type Pauli operators. General disjoint Pauli operators can be reduced to the $Z$-type case through local Clifford gates. We write $U:=\bigcup_{a=1}^mS_a$ for the union of the target supports.

In each short evolution interval, apply a Pauli pulse
\begin{equation}
    R(\boldsymbol x,\boldsymbol z)
    :=\prod_{j=1}^N X_j^{x_j}Z_j^{z_j},
    \label{eq:main-correlated-reshaping-pulse}
\end{equation}
where all $z_j$ are independent uniformly random bits. The bits $x_j$ are independent and uniformly random for $j\notin U$, while on each target support $S_a$ they are sampled uniformly subject to the even-parity constraint
\begin{equation}
    \bigoplus_{j\in S_a}x_j=0,\qquad\forall a.
    \label{eq:main-reshaping-parity}
\end{equation}
The constraints on different target supports are sampled independently.

To see the effect of this average, consider a $Z$-type Pauli string $Z_T$. Its conjugation sign is
\begin{equation}
    R(\boldsymbol x,\boldsymbol z)Z_TR(\boldsymbol x,\boldsymbol z)^\dagger
    =(-1)^{\sum_{j\in T}x_j}Z_T.
    \label{eq:main-Z-string-sign}
\end{equation}
The parity constraint in Eq.~\eqref{eq:main-reshaping-parity} preserves the complete target string $Z_{S_a}$, whereas every proper nonempty substring of $Z_{S_a}$ acquires a plus or a minus sign with equal probability and averages to zero. The independent choices of $x_j$ outside $U$ remove terms acting outside the target supports, while the independent $z_j$ remove all terms containing $X$ or $Y$. Consequently, assuming that no tensor product of two or more distinct $P_a$ appears in the original Hamiltonian, the effective Hamiltonian after reshaping is 
\begin{equation}
    H_{\mathrm{eff}}=\sum_{a=1}^m\lambda_a P_{a}.
    \label{eq:main-disjoint-reshaped-Hamiltonian}
\end{equation}
All selected coefficients are retained with their original values, while the other terms are averaged away. The condition that no tensor product of distinct  $P_{a}$ appears in the original Hamiltonian is crucial. These products also commute with all sampled $R(x,z)$ operators and are therefore preserved under reshaping. For two-body Hamiltonians considered in this work, this condition can always be satisfied.  

\subsection{Robust phase estimation}
\label{subsec:rpe-main}

Once an experiment produces an oscillation whose frequency is an unknown coefficient, learning that coefficient becomes a phase-estimation problem. Consider first a single qubit evolving under $\lambda Z$. Starting from the initial state $\ket+=(\ket0+\ket1)/\sqrt2$, evolve under $\lambda Z$ for time $t$. Then separate measurements of $X$ and $Y$ give
\begin{equation}
    \langle X\rangle_t=\cos(2\lambda t),
    \qquad
    \langle Y\rangle_t=\sin(2\lambda t).
    \label{eq:rpe-ramsey-main}
\end{equation}
These two expectations determine the phase $2\lambda t$ modulo $2\pi$. Longer evolution improves sensitivity to $\lambda$, but also introduces ambiguity because several possible values can produce the same phase. Multiscale phase estimation resolves this ambiguity using progressively longer experiments~\cite{Higgins2009,KimmelLowYoder2015}. 

For $|\lambda|\leq\Lambda$, choose $t_0=\pi/(4\Lambda)$ and $t_k=2^kt_0$. The initial phase lies in $[-\pi/2,\pi/2]$ and can be estimated without ambiguity. At each later scale, infer a phase $\widehat\varphi_k\in(-\pi,\pi]$ from the two measured quadratures and select the value
\begin{equation}
    \widehat\lambda_k
    =\underset{\lambda'\in\{(\widehat\varphi_k+2\pi\ell)/(2t_k):\,\ell\in\mathbb Z\}}
      {\operatorname{argmin}}
      |\lambda'-\widehat\lambda_{k-1}|,
    \qquad
    \widehat\lambda_0=\frac{\widehat\varphi_0}{2t_0}.
    \label{eq:rpe-unwrapping-main}
\end{equation}
If each phase estimate has error at most $\gamma<\pi/3$ modulo $2\pi$, this rule selects the correct branch and gives $|\widehat\lambda_k-\lambda|\leq\gamma/(2t_k)$. Hence a longest evolution time $t_K=\Theta(1/\varepsilon)$ suffices for accuracy $\varepsilon$.

\emph{Robust} phase estimation allows small, unknown additive deviations in the measurement probabilities~\cite{KimmelLowYoder2015,KimmelLowYoder2021Erratum}. Write the probabilities of the $+1$ outcomes as
\begin{equation}
    \widetilde p_X(t)=\frac{1+\cos(2\lambda t)}{2}+b_X(t),
    \qquad
    \widetilde p_Y(t)=\frac{1+\sin(2\lambda t)}{2}+b_Y(t).
    \label{eq:rpe-biased-probabilities}
\end{equation}
A sufficiently small uniform bound on $|b_X(t_k)|$ and $|b_Y(t_k)|$ ensures constant phase accuracy after repeated sampling. The biases need not shrink with $\varepsilon$: a constant phase uncertainty becomes a coefficient uncertainty of order $1/t_k$. The bound must, however, hold through the longest time used. 

At each scale $k$, repeat the experiment $R_k$ times for each measurement basis to estimate the two outcome probabilities. Provided the systematic biases remain below a sufficiently small constant, estimating these probabilities to a fixed additive accuracy keeps the phase error below $\gamma$. Concentration bounds imply that $R_k=O(\log(1/\delta_k))$ repetitions per basis suffice to achieve this accuracy with failure probability at most $\delta_k$. For a target failure probability $0<\delta\leq1/2$, choose
\begin{equation}
    \delta_k:=\delta\,2^{k-K-1},
    \qquad
    R_k=\left\lceil C\left[\log\frac{1}{\delta}+K-k\right]\right\rceil,
    \label{eq:rpe-repetitions-main}
\end{equation}
where $C$ is a sufficiently large constant determined by the allowed bias and phase-error threshold. Since $\sum_{k=0}^{K}\delta_k<\delta$, all phase estimates satisfy the required accuracy with probability at least $1-\delta$. This allocation uses more repetitions at shorter evolution times, where they cost less. Using $t_k=2^{k-K}t_K$, the total evolution time is
\begin{equation}
    T_{\rm phase}=2\sum_{k=0}^{K}R_kt_k
    =O\!\left(t_K\log\frac{1}{\delta}\right)
    =O\!\left(\frac{1}{\varepsilon}\log\frac{1}{\delta}\right).
    \label{eq:rpe-time-main}
\end{equation}
The factor of two accounts for the two measurement bases. At fixed confidence, the total signal-evolution time scales as $1/\varepsilon$, giving the Heisenberg scaling used below. For simultaneous estimation of $L$ coefficients, allocating failure probability $\delta/L$ to each replaces $\log(1/\delta)$ by $\log(L/\delta)$. 

\section{The fundamental lower bound}
\label{subsec:benchmark-scalings}
Two standard limitations provide useful benchmarks for this learning problem. We state them for fixed success probability greater than $1/2$ and nontrivial accuracy $\varepsilon<c\Lambda$, with a sufficiently small fixed $c>0$.

First, consider two allowed Hamiltonians differing in a single Pauli coefficient by $\Delta=\Theta(\varepsilon)$, with disjoint accuracy intervals. Their generator difference has operator norm $|\Delta|$. Over total evolution time $T_{\rm tot}$, the trace distance between the final states generated under the two hypotheses is at most of order $|\Delta|T_{\rm tot}$, even with interleaved parameter-independent controls. Successful estimation must distinguish these two cases with constant advantage and therefore requires
\begin{equation}
    T_{\rm tot}=\Omega(1/\varepsilon).
    \label{eq:heisenberg-benchmark}
\end{equation}
This is the usual Heisenberg-limit benchmark~\cite{Giovannetti2006}.

Second, a change to a coefficient near the far endpoint must influence the bright qubit before it can be detected there. For nearest-neighbor interactions of bounded strength, Lieb--Robinson bounds limit this propagation to a speed of order $\Lambda$~\cite{LiebRobinson1972,NachtergaeleSims2010}. The distance is $\Theta(N)$, giving the worst-case locality benchmark
\begin{equation}
    T_{\rm tot}=\Omega(N/\Lambda).
    \label{eq:locality-benchmark}
\end{equation}
The influence outside the corresponding propagation region is exponentially small, so combining many short experiments does not remove this bound on total evolution time. This locality constraint continues to hold even in the presence of arbitrary time-dependent single-qubit controls, since on-site terms do not increase the Lieb--Robinson propagation velocity set by the inter-site interactions~\cite{LiebRobinson1972}.

\section{Parallel learning}\label{sec:robust}
We now construct a protocol that learns many Hamiltonian coefficients in parallel using measurements and resets at the single bright endpoint. The central challenge is to prepare and probe the dark qubits through interactions whose strengths are themselves unknown. We first show how quantum signal processing enables approximate SWAP operations without first estimating those interaction strengths. These operations allow quantum information to move throughout the chain.

We then use these approximate SWAP operations both to initialize the register and to bring information to the bright qubit for measurement. Repeated resets of the bright qubit, combined with exchanges of neighboring states, prepare the entire chain approximately in the all-zero state. Hamiltonian reshaping then isolates a set of terms acting on disjoint qubits or pairs, allowing their coefficients to be encoded simultaneously in the phases of separate qubit states. These states are transferred to the bright qubit and measured one at a time. Repeating this procedure for different evolution times and sets of terms yields estimates of all Hamiltonian coefficients.

\subsection{Robust SWAP construction}
We first explain how to construct robust SWAP gates for Hamiltonian bonds satisfying the reference-backbone assumption in  Eq.~\eqref{eq:basic-transport-assumption}. For simplicity, we first consider the two-qubit example:
\begin{equation}
    H_J = J Z_1Z_2,
    \qquad
    J_- \leq |J| \leq J_+,
    \label{eq:single_bond}
\end{equation}
where $J$ is the unknown coefficient, while the interval endpoints $J_-, J_+ > 0$ are known.  All control sequences constructed below depend only on $J_-$, $J_+$, and the target accuracy, and are independent of the value and sign of $J$. The unknown coupling prevents us from choosing a calibrated evolution time for a standard SWAP decomposition. To construct a robust alternative, we introduce the following encoded Pauli operators.
\begin{definition}[Encoded Pauli operators]
\label{def:encoded_paulis}
    Define the encoded Pauli operators
\begin{equation}
    \overline{X}=Z_1Z_2,
    \qquad
    \overline{Y}=X_1Z_2,
    \qquad
    \overline{Z}=Y_1I_2 .
    \label{eq:encoded_paulis}
\end{equation}
These operators satisfy the Pauli algebra
\begin{equation}
    \overline{X}^2 = \overline{Y}^2 = \overline{Z}^2 = I,
    \qquad
    \overline{X}\overline{Y} = i\overline{Z}.
\end{equation}
\end{definition}
For the encoded qubit, the interaction term $JZ_1Z_2$ acts as an effective $X$-rotation with an unknown rotation angle determined by $J$, while the controllable single-qubit rotation generated by $Y_1$ implements an encoded $Z$-phase rotation. Alternating these two operations realizes the standard quantum signal processing (QSP) structure~\cite{Low2016Methodology} in this encoded Pauli algebra. The construction acts on the full two-qubit Hilbert space and requires no restriction on the input state.

For a free-evolution interval of duration $\tau$, the encoded signal query is
\begin{equation}
        e^{-iH_J\tau}= e^{-i\theta_J\overline{X}/2},
\end{equation}
where the accumulated angle is $\theta_J=2J\tau$. Conjugating this evolution by a single-qubit rotation about $Y_1$ produces a family of phase-controlled QSP signal operators,
\begin{equation}
    R_{\phi}(\theta_J) = e^{-i\phi Y_1/2} e^{-iH_J\tau} e^{+i\phi Y_1/2}.
    \label{eq:primitive_rotation}
\end{equation}
By Definition \ref{def:encoded_paulis}, each QSP signal operator has the form
\begin{equation}
    R_{\phi}(\theta_J) = \exp\!\left[ -\frac{i\theta_J}{2} \left( \cos\phi\,\overline{X} + \sin\phi\,\overline{Y} \right) \right].
    \label{eq:phase_controlled_signal_rotation}
\end{equation}
This is the primitive rotation of the equiangular composite-gate
formalism of Ref.~\cite{Low2016Methodology}. 

For a general signal angle $\theta$, write
$S(\theta):=e^{-i\theta\overline X/2}$ and
$Z(\phi):=e^{-i\phi\overline Z/2}$, so that
$R_\phi(\theta):=Z(\phi)S(\theta)Z(-\phi)$ agrees with
Eq.~\eqref{eq:phase_controlled_signal_rotation}.

\begin{lemma}[Equiangular quantum transformation]
\label{lem:equiangular_response}
Let $L\geq1$ be the number of primitive rotations, equivalently the number of signal queries, and let $\boldsymbol\phi=(\phi_1,\ldots,\phi_L)$.  The length-$L$ sequence
\begin{equation}
    U_{\boldsymbol\phi}(\theta):=R_{\phi_L}(\theta)\cdots R_{\phi_1}(\theta)
    \label{eq:equiangular_sequence}
\end{equation}
has the polynomial representation
\begin{equation}
    U_{\boldsymbol\phi}(\theta) = A_L(x)I+iB_L(x)\overline Z
    +i\kappa_L(x)
      \bigl[C_L(y)\overline X+D_L(y)\overline Y\bigr],
    \label{eq:quantum_response}
\end{equation}
where $x=\cos(\theta/2)$, $y=\sin(\theta/2)$, and
$\kappa_L(x)=1$ for odd $L$, while $\kappa_L(x)=x$ for even $L$.
The four real polynomials have degree at most $L$; all are odd when $L$
is odd, whereas $A_L,B_L$ are even and $C_L,D_L$ are odd when $L$ is
even.  Conversely, polynomials with these degree and parity properties
are achievable precisely when
\begin{equation}
    A_L(1)=1,\qquad B_L(1)=0,\qquad
    A_L(x)^2+B_L(x)^2
    +\kappa_L(x)^2\bigl[C_L(y)^2+D_L(y)^2\bigr]=1
    \quad (x^2+y^2=1).
    \label{eq:response_achievability}
\end{equation}
The phases $\boldsymbol\phi$ can be computed efficiently from any achievable tuple.
\end{lemma}

\begin{proof}
The result follows from Theorem~1 and Lemma~1 of Ref.~\cite{Low2016Methodology}. Theorem~1 characterizes the polynomial tuples achievable by an equiangular composite sequence, while Lemma~1 establishes that any such achievable tuple can be implemented using exactly $L$ phases, which can be computed in time $\operatorname{poly}(L)$. To apply these results here, we replace the standard Pauli operators $(X,Y,Z)$ in Ref.~\cite{Low2016Methodology} by the encoded operators $(\overline X,\overline Y,\overline Z)$.
\end{proof}

By collecting adjacent encoded-$Z$ rotations,
Eq.~\eqref{eq:equiangular_sequence} can equivalently be written as an interleaved signal-and-phase sequence, namely, the QSP sequence \cite{Dong_2021, LowChuang2019} (up to phase convention). By specifically designing the polynomial tuples to approximate the sign function, we can construct the uniform robust entangling rotation.

For the remainder of the construction, choose
$\tau=\pi/[2(J_-+J_+)]$ and define
$\Delta:=\sin[\pi J_-/(J_-+J_+)]$.  Then
$\theta_J=\pi J/(J_-+J_+)$ obeys
$|\sin\theta_J|\geq\Delta$, so the response polynomial only needs to
approximate the sign function away from the interval $(-\Delta,\Delta)$.

\begin{lemma}[Uniform robust entangling rotation]
\label{lem:robust_entangler}
For $H_J=JZ_1Z_2$ with $J_-\leq|J|\leq J_+$, and for every fixed
$\alpha\in[-\pi,\pi]$ and $0<\eta<1$, there exists a $J$-independent
phase sequence $\boldsymbol\phi$ such that
$U_\alpha(J):=U_{\boldsymbol\phi}(\theta_J)$ satisfies
\begin{equation}
    \sup_{J_-\leq |J|\leq J_+}
    \left\|
        U_{\alpha}(J) - e^{-i\,\operatorname{sgn}(J)\alpha\overline X}
    \right\|
    \leq \eta .
    \label{eq:robust_entangler_bound}
\end{equation}
The sequence length and total interaction time are
$L=O[\Delta^{-1}\log(1/\eta)]$ and
$T_{\mathrm{ent}}(\eta)=O[J_-^{-1}\log(1/\eta)]$, respectively.
\end{lemma}

\begin{proof}[Proof sketch]
An odd bounded polynomial approximates $\operatorname{sgn}(s)$ uniformly
on $|s|\geq\Delta$ with degree $O[\Delta^{-1}\log(1/\eta)]$.
Lemma~\ref{lem:equiangular_response} converts this polynomial into a
phase sequence whose response approximates
$e^{-i\operatorname{sgn}(J)\alpha\overline X}$.  Multiplying the sequence
length by the query duration $\tau$ gives the stated interaction time.
Supplementary Sec.~\ref{appendix:robust_swap} supplies the explicit response
polynomials and error conversion.
\end{proof}

The SWAP construction then follows from the identity $XX+YY+ZZ=2\mathrm{SWAP}-I$. 

\begin{theorem}[Parameter-independent robust SWAP]
\label{thm:robust_swap}
For the bond in Eq.~\eqref{eq:single_bond} and every $0<\eta<1$, there
exists a $J$-independent control sequence.  If
$\widetilde{\mathcal S}_J$ and $\mathcal S$ denote the induced approximate
and ideal SWAP channels, then
\begin{equation}
 \sup_{J_-\leq|J|\leq J_+}
 \|\widetilde{\mathcal S}_J-\mathcal S\|_\diamond\leq\eta.
    \label{eq:robust_swap_error}
\end{equation}
The required interaction time is
$T_{\mathrm{SWAP}}(\eta)=O[J_-^{-1}\log(1/\eta)]$.
\end{theorem}

\begin{proof}[Proof sketch]
Apply Lemma~\ref{lem:robust_entangler} with $\alpha=\pi/4$ and error
$\eta/6$, and use local Clifford conjugations to obtain the corresponding
$XX$, $YY$, and $ZZ$ rotations.  They commute and satisfy
$XX+YY+ZZ=2\operatorname{SWAP}-I$, so their ideal product is SWAP up to
a global phase.  A three-term telescoping bound gives operator-norm error
at most $\eta/2$; the inequality
$\|\mathcal U-\mathcal V\|_\diamond\leq2\|U-V\|$ then gives channel error
at most $\eta$.
Only three robust rotations are used, so the interaction-time scaling is
unchanged.
\end{proof}

We now describe how to extend the bond-wise construction to the entire reference backbone. Let $E_o$ and $E_e$ denote the bonds $(i,i+1)$ with odd and even $i$, respectively. For every $1\leq i<N$, the reference-backbone assumption specifies $P_i^{\mathrm{ref}}=\sigma_i^{a_i}\sigma_{i+1}^{b_i}$ and $g_i=J_{i,i+1}^{a_ib_i}$ satisfying $J_-\leq|g_i|\leq J_+$. Define the reference-component Hamiltonians by
\begin{equation}
     H_{{\rm ref},c}=\sum_{(i,i+1)\in E_c}g_iP_i^{\rm ref},\qquad c\in\{e,o\}.
    \label{eq:matching-reference-Hamiltonian}
\end{equation}
For fixed $c\in\{o,e\}$, the operators $P_i^{\mathrm{ref}}$ in $H_{\mathrm{ref},c}$ have disjoint qubit supports. We can therefore apply the Hamiltonian reshaping technique developed in Sec.~\ref{subsec:reshaping-main} to obtain $H_{\mathrm{ref},c}$ as the effective Hamiltonian.

For each $a\in\{X,Y,Z\}$, choose a local Clifford gate $C_a$ satisfying $C_a\sigma^aC_a^\dagger=Z$. Conjugating the effective Hamiltonian by these gates gives
\begin{equation}\label{eq:local-Clifford}
    \Bigl(\otimes_{(i,i+1)\in E_{c}}(C_{a_{i}}\otimes C_{b_{i}})\Bigr)H_{\mathrm{ref},c}\Bigl(\otimes_{(i,i+1)\in E_{c}}(C_{a_{i}}^{\dagger}\otimes C_{b_{i}}^{\dagger})\Bigr)=\sum_{(i,i+1)\in E_c}g_iZ_iZ_{i+1}.
\end{equation}
The same QSP phase sequence can then be applied on every bond simultaneously. We denote the two implemented layers by
\begin{equation}
 \widetilde S_c:=
 \prod_{(i,i+1)\in E_c}
 \widetilde{\operatorname{SWAP}}_{i,i+1},
 \qquad c\in\{o,e\},
 \label{eq:robust-swap-layers}
\end{equation}
where each factor is the unitary produced by Theorem~\ref{thm:robust_swap}.

\subsection{Parallel learning architecture}
The SWAP construction allows the bright qubit to serve both as a source of prepared states and as the measurement site for the entire chain. We first use it to initialize the register. The reset channel at the bright qubit at site $1$ can be written as
\begin{equation}
    \mathcal R_1(\rho):=\ketbra{0}{0}_1\otimes\operatorname{Tr}_1(\rho).
    \label{eq:probe-reset}
\end{equation}
Combining this reset with the robust SWAP layers in Eq.~\eqref{eq:robust-swap-layers} gives the following preparation procedure, which requires no knowledge of the initial register state.

\begin{mybox}[label=box:initialization]{Register initialization through the bright qubit}
\begin{enumerate}
    \item Reset the bright qubit to $\ket0$.
    \item Repeat the following cycle $N-1$ times: apply the even-bond robust SWAP layer $\widetilde S_e$, then the odd-bond robust SWAP layer $\widetilde S_o$, and reset the bright qubit to $\ket{0}$ again.
    \item Use the resulting register state, denoted by $\widetilde\rho_0$, as the input to the learning procedure.
\end{enumerate}
\end{mybox}

To see why this prepares the whole chain, replace the approximate layers by exact SWAP layers $S_e$ and $S_o$. The combined permutation $D:=S_oS_e$ cycles through all $N$ sites. Repeating it brings each original qubit subsystem to the bright site, where it is reset. After the initial reset and $N-1$ cycles, every subsystem has therefore been replaced by $\ket0$. This remains true for an initially entangled register since SWAPs exchange subsystems, and each reset discards the arriving subsystem and its correlations with the rest.

Let $\mathcal D$ denote the channel induced by $D$, and let $\widetilde{\mathcal D}$ denote the channel induced by $\widetilde D:=\widetilde S_o\widetilde S_e$. The ideal initialization satisfies
\begin{equation}
    \mathcal I:=(\mathcal R_1\circ\mathcal D)^{N-1}\circ\mathcal R_1,
    \qquad \mathcal I(\rho)=\ketbra{0}{0}^{\otimes N},
    \label{eq:ideal-probe-initialization}
\end{equation}
for every initial state $\rho$, whereas Box~\ref{box:initialization} implements
\begin{equation}
    \widetilde{\mathcal I}:=(\mathcal R_1\circ\widetilde{\mathcal D})^{N-1}\circ\mathcal R_1,
    \label{eq:approximate-probe-initialization}
\end{equation}
The procedure uses $N$ bright-qubit resets and $2(N-1)$ parallel SWAP layers, containing $(N-1)^2$ two-qubit SWAPs in total. Its output is $\widetilde\rho_0:=\widetilde{\mathcal I}(\rho)$. Section~\ref{subsec:parallel-error-analysis} bounds its trace-norm distance from $\ketbra{0}{0}^{\otimes N}$ by $(N-1)^2\eta$, where $\eta$ bounds the diamond-norm error of each SWAP channel. 

We next combine this initialization procedure with parallel encoding and readout. Box~\ref{box:parallel} specifies the implemented protocol using approximate SWAPs. We explain its operation using exact SWAPs and account for both initialization and readout errors in the next subsection.

For clarity, we give the site assignments for even $N$; for odd $N$, the same protocol applies with the corresponding endpoint counts adjusted. Let $V_o$ and $V_e$ be the odd and even sites, and retain the corresponding bond sets $E_o$ and $E_e$. We divide the coefficients into six one-body Hamiltonians,
\begin{equation}\label{eq:one-body}
    \mathfrak G_1=\left\{H_c^\alpha=\sum_{i\in V_c}h_i^\alpha\sigma_i^\alpha:
    c\in\{o,e\},\ \alpha\in\{X,Y,Z\}\right\},
\end{equation}
and eighteen two-body Hamiltonians,
\begin{equation}\label{eq:two-body}
    \mathfrak G_2=\left\{H_c^{\alpha\beta}=\sum_{(i,i+1)\in E_c}J_{i,i+1}^{\alpha\beta}\sigma_i^\alpha\sigma_{i+1}^\beta:
    c\in\{o,e\},\ \alpha,\beta\in\{X,Y,Z\}\right\}.
\end{equation}
The terms in each $A\in\mathfrak G:=\mathfrak G_1\cup\mathfrak G_2$ have disjoint supports. The one-body groups also contain no adjacent sites, so all groups satisfy the isolation condition in Sec.~\ref{subsec:reshaping-main}. Write $A=\sum_{j\in I_A}\lambda_jP_j$, where $I_A$ labels the selected terms. For each term, choose one \emph{record qubit} $r_j$, whose relative phase will carry information about $\lambda_j$. We call the resulting qubit state a \emph{phase record}. For a two-body term, the other qubit $q_j$ is the \emph{anchor}; its target state $\ket0$ allows the interaction to generate a phase on the record qubit. We use the following assignments:
\begin{center}
\renewcommand{\arraystretch}{1.2}
\begin{tabular}{c|c|c|c}
    selected $A$ & $I_A$ & coefficient $\lambda_j$ & site assignment \\ \hline
    $H_o^\alpha$ & $1\leq j\leq N/2$ & $h_{2j-1}^\alpha$ & $r_j=2j-1$ \\
    $H_e^\alpha$ & $1\leq j\leq N/2$ & $h_{2j}^\alpha$ & $r_j=2j$ \\
    $H_o^{\alpha\beta}$ & $1\leq j\leq N/2$ & $J_{2j-1,2j}^{\alpha\beta}$ & $r_j=2j,\ q_j=2j-1$ \\
    $H_e^{\alpha\beta}$ & $1\leq j\leq N/2-1$ & $J_{2j,2j+1}^{\alpha\beta}$ & $r_j=2j,\ q_j=2j+1$
\end{tabular}
\end{center}
Conjugating the reshaped evolution by known local Clifford gates maps the selected Pauli operators to $Z$ operators, giving the effective Hamiltonian
\begin{equation}
    A_Z=
    \begin{cases}
        \displaystyle\sum_{j\in I_A}\lambda_j Z_{r_j},&A\in\mathfrak G_1,\\[2mm]
        \displaystyle\sum_{j\in I_A}\lambda_j Z_{q_j}Z_{r_j},&A\in\mathfrak G_2.
    \end{cases}
    \label{eq:parallel-selected-Z-Hamiltonian}
\end{equation}
Box~\ref{box:parallel} combines this encoding with initialization and sequential readout through the bright qubit.

\begin{mybox}[label=box:parallel]{Parallel learning through the bright qubit}
\begin{enumerate}
    \item \emph{Select.} Choose $A\in\mathfrak G$, a measurement basis $b\in\{X,Y\}$, and an evolution time $t_k$ from the Ramsey schedule
    \begin{equation*}
        t_0:=\frac{\pi}{4\Lambda},\qquad t_k:=2^kt_0,\qquad
        K:=\min\left\{q\in\mathbb Z_{\geq0}:2^qt_0\geq\frac{\pi}{12\varepsilon}\right\},
        \quad k=0,\ldots,K.
    \end{equation*}
    \item \emph{Reshape.} Choose the Hamiltonian reshaping sequence and local Clifford transformations that realize $A_Z$ in Eq.~\eqref{eq:parallel-selected-Z-Hamiltonian}.
    \item\label{step:encode} \emph{Initialize and encode.} Prepare $\widetilde\rho_0$ using Box~\ref{box:initialization}. Apply a Hadamard gate to each record qubit $r_j$, then evolve under $A_Z$ for time $t_k$.
    \item \emph{Align.} If $A=H_o^\alpha$, no alignment is needed. Otherwise, apply the odd-bond layer $\widetilde S_o$.
    \item\label{step:drain} \emph{Drain.} For $m=1,\ldots,|I_A|$, measure the bright qubit in basis $b$ and reset it to $\ket0$. If unmeasured records remain, apply $\widetilde S_e$ followed by $\widetilde S_o$. Associate the $m$-th outcome with the $m$-th coefficient in the site assignment above.
    \item \emph{Estimate.} For each $A$, $k$, and $b$, repeat steps~\ref{step:encode}--\ref{step:drain}
    \begin{equation*}
        R_k=\left\lceil C\left[\log\!\frac{2(12N-9)}{\delta}+K-k\right]\right\rceil
    \end{equation*}
    times, with a sufficiently large constant $C$. Denote the outcome for coefficient $\lambda_j$ in repetition $\ell$ by $x_{j,k,\ell}^{(b)}\in\{+1,-1\}$, and form
    \begin{equation*}
        \widehat c_{j,k}:=\frac{1}{R_k}\sum_{\ell=1}^{R_k}x_{j,k,\ell}^{(X)},\qquad
        \widehat s_{j,k}:=\frac{1}{R_k}\sum_{\ell=1}^{R_k}x_{j,k,\ell}^{(Y)}.
    \end{equation*}
    Apply the robust phase estimator of Sec.~\ref{subsec:rpe-main} to these averages across $t_0,\ldots,t_K$ to estimate each $\lambda_j$.
\end{enumerate}
\end{mybox}

To explain the protocol, first consider exact SWAPs, for which Box~\ref{box:initialization} prepares $\ket0^{\otimes N}$ exactly.
Let $R_A:=\{r_j:j\in I_A\}$ denote the set of record sites. Applying the Hadamard gates on these sites gives
\begin{equation}
    \ket{\Psi_A}:=\left(\prod_{j\in I_A}\mathrm{Had}_{r_j}\right)\ket0^{\otimes N}
    =\bigotimes_{j\in I_A}\ket+_{r_j}\otimes\ket0_{[N]\setminus R_A}.
    \label{eq:parallel-target-input}
\end{equation}
For a one-body term, $A_Z$ directly rotates the record qubit about $Z$. For a two-body term, the anchor remains in $\ket0$, so $Z_{q_j}\ket0_{q_j}=\ket0_{q_j}$ gives the same rotation on the record. Since the selected terms have disjoint supports, evolution for time $t_k$ produces
\begin{equation}
    \ket{\Psi_A(t_k)}=e^{-iA_Zt_k}\ket{\Psi_A}
    =\bigotimes_{j\in I_A}\left(e^{-i\lambda_jt_kZ}\ket+\right)_{r_j}
    \otimes\ket0_{[N]\setminus R_A}.
    \label{eq:parallel-record-factorization}
\end{equation}
Thus one evolution encodes all coefficients in $A$ in separate qubit states.

For $A=H_o^\alpha$, these states already occupy the odd sites. In the other cases, the alignment layer moves each record from site $2j$ to site $2j-1$. The first record is then at the bright qubit. After it is measured and the bright qubit is reset, the even-bond layer followed by the odd-bond layer moves the next record along $3\to2\to1$ and advances each later record by two sites. Repeating this cycle reads all records in spatial order. Exact SWAPs only exchange qubit states, so they preserve the encoded phases.

When the record for $\lambda_j$ reaches the bright qubit, its state is still $e^{-i\lambda_jt_kZ}\ket+$. The two measurement bases therefore give
\begin{equation}
    \mathbb E\!\left[x_{j,k}^{(X)}\right]=\cos(2\lambda_jt_k),\qquad
    \mathbb E\!\left[x_{j,k}^{(Y)}\right]=\sin(2\lambda_jt_k).
    \label{eq:parallel-family-signals}
\end{equation}
Measurements at the prescribed evolution times supply the Ramsey signals needed to recover each coefficient using Sec.~\ref{subsec:rpe-main}. Repeating the procedure over $\mathfrak G$ covers all coefficients of the Hamiltonian. In the implemented protocol, approximate SWAPs affect both the prepared state and its subsequent transfer. The next subsection bounds the resulting changes in the measured signals and shows that the same estimation procedure remains valid.

\subsection{Error analysis}
\label{subsec:parallel-error-analysis}

We now determine the accuracy required of the robust SWAP gates and bound the total time needed to learn the Hamiltonian to precision $\varepsilon$.

We first bound the initialization error. For a bond $e$, let $W_e=\operatorname{SWAP}_e$ and $\widetilde W_e$ be the implemented unitary. Their channels are $\mathcal W_e(\rho):=W_e\rho W_e^\dagger$ and $\widetilde{\mathcal W}_e(\rho):=\widetilde W_e\rho\widetilde W_e^\dagger$. We call the implemented SWAP $\eta$-approximate if
\begin{equation}
    \|\widetilde{\mathcal W}_e-\mathcal W_e\|_\diamond\leq\eta.
    \label{eq:individual-swap-channel-error}
\end{equation}
The initialization circuit contains $(N-1)^2$ SWAPs. Repeated use of the triangle inequality, together with contractivity under channels, gives
\begin{equation}
    \|\widetilde\rho_0-\ketbra{0}{0}^{\otimes N}\|_1\leq(N-1)^2\eta
    \label{eq:initialization-trace-error}
\end{equation}
for every initial register state $\rho$.

When SWAPs are exact, different readout rounds reveal coefficients on disjoint supports and their outcomes are independent. Approximate SWAPs can correlate these outcomes, but reconstructing a given coefficient requires only the marginal distribution of the corresponding readout round. It therefore suffices to control each marginal rather than the full joint distribution.

\begin{lemma}\label{lem:robust-marginal-distribution}
    Let $p_m$ and $\widetilde p_m$ be the marginal outcome distributions in measurement of round $m$ for the exact-$\mathrm{SWAP}$ and $\eta$-approximate $\widetilde{\mathrm{SWAP}}$ circuits, respectively. Then
    \begin{equation}
        d_{\mathrm{TV}}(p_m,\widetilde p_m)\leq (N-1)^2\eta+2mN\eta,
    \end{equation}
    where $d_{\mathrm{TV}}$ is the total variation distance between two probability distributions.
\end{lemma}

\begin{proof}[Proof sketch]
Let $\rho_{\rm in}$ and $\rho_{\rm in}^{\rm ideal}$ be the implemented and ideal encoded states immediately before alignment. The same Hadamard gates and encoding evolution act on $\widetilde\rho_0$ and $\ketbra0{0}^{\otimes N}$, so unitary invariance gives
\begin{equation}
    \|\rho_{\rm in}-\rho_{\rm in}^{\rm ideal}\|_1\leq(N-1)^2\eta.
\end{equation}
Let $\widetilde{\mathcal C}_m$ and $\mathcal C_m$ be the implemented and ideal channels from these encoded states through measurement round $m$, including alignment, the preceding SWAP layers, and measurement and reset with outcomes stored in classical registers. At most $2m$ SWAP layers occur before this measurement, and each layer contains at most $N/2$ SWAPs. The triangle inequality therefore gives $\|\widetilde{\mathcal C}_m-\mathcal C_m\|_\diamond\leq mN\eta$. Combining the input-state and channel errors gives
\begin{equation}
    \|\widetilde{\mathcal C}_m(\rho_{\rm in})-\mathcal C_m(\rho_{\rm in}^{\rm ideal})\|_1
    \leq(N-1)^2\eta+mN\eta.
\end{equation}
Tracing out every register except the classical outcome of round $m$ cannot increase this norm. For classical states, the trace-norm distance is twice the total variation distance. This proves the stated, conservative bound. Supplementary Sec.~\ref{app:marginal-lemma} gives the measurement-and-reset channel and the detailed comparison.
\end{proof}

Each marginal distribution determines one of the two Ramsey quadratures associated with a single Hamiltonian coefficient. If an empirical distribution $\widehat p_m$ estimates $\widetilde p_m$ to sampling error $d_{\mathrm{TV}}(\widehat p_m,\widetilde p_m)\leq\alpha$, then Lemma~\ref{lem:robust-marginal-distribution} and the triangle inequality give
\begin{equation}
    d_{\mathrm{TV}}(\widehat p_m,p_m)\leq\alpha+(N-1)^2\eta+2mN\eta.
    \label{eq:robust-marginal-total-error}
\end{equation}
Thus it is enough to keep the sampling error and the error of each marginal below a constant threshold.  The robust phase estimator introduced in Sec.~\ref{subsec:rpe-main} can then recover every coefficient with Heisenberg-limited precision dependence.

\begin{theorem}[Robust parallel Hamiltonian learning on a chain]\label{thm:robust-parallel-learning}
Consider the setting of Sec.~\ref{subsec:Hamiltonian-model} with the reference couplings satisfying Eq.~\eqref{eq:basic-transport-assumption}.  For any $0<\varepsilon<\Lambda$ and $0<\delta<1$, the robust parallel-learning protocol produces an estimator $\widehat{\boldsymbol\lambda}$ such that
\begin{equation}
    \Pr\!\left(\left\|\widehat{\boldsymbol\lambda}-\boldsymbol\lambda\right\|_\infty\leq\varepsilon\right)\geq 1-\delta.
    \label{eq:robust-learning-guarantee}
\end{equation}
It suffices to implement every nearest-neighbor $\widetilde{\mathrm{SWAP}}$ with diamond-norm error $\eta=O(N^{-2})$. The total native-Hamiltonian evolution time, including initialization in every repetition, is
\begin{equation}
\begin{split}
    T_{\mathrm{tot}}=O\!\left[\frac{1}{\varepsilon}\log\!\left(\frac{N}{\delta}\right)
    +\frac{N}{J_-}\log N
    \left(\log\!\left(\frac{N}{\delta}\right)+\log\!\left(\frac{2\Lambda}{\varepsilon}\right)\right)
    \log\!\left(\frac{2\Lambda}{\varepsilon}\right)\right].
    \label{eq:robust-learning-time}
\end{split}
\end{equation}
In particular, for fixed $\varepsilon$, $\delta$, $\Lambda$, and $J_\pm$, the total time is $O(N\log^2N)=\widetilde O(N)$, while its leading precision dependence retains Heisenberg scaling.
\end{theorem}

\begin{proof}[Proof sketch]
We give the main steps here; the detailed concentration and summation arguments appear in Supplementary Sec.~\ref{app:robust-learning-proof}. Choose the diamond-norm error of each robust SWAP to be $\eta=c/N^2$ for a sufficiently small constant $c$. For every readout round $m=O(N)$, Lemma~\ref{lem:robust-marginal-distribution} then gives $d_{\mathrm{TV}}(p_m,\widetilde p_m)=O(c)$. Then estimating each empirical marginal to a sufficiently small constant sampling error still determines its Ramsey quadrature to constant accuracy. By Theorem~\ref{thm:robust_swap}, a SWAP with $\eta=O(N^{-2})$ can be implemented in logarithmic time $O(J_-^{-1}\log N)$.

We next convert these constant-accuracy Ramsey signals into coefficient estimates. For each coefficient $\lambda_a$, the $X$- and $Y$-basis measurements at time $t_k$ determine the phase $2\lambda_a t_k$ modulo $2\pi$.  Set $t_0=\pi/(4\Lambda)$ and $t_k=2^kt_0$.  The initial phase is unambiguous because $|\lambda_a|\leq\Lambda$; at every later scale the preceding estimate selects the unique compatible phase branch.  Constant phase accuracy at each scale gives coefficient error $O(1/t_k)$, so $t_K=\Theta(1/\varepsilon)$ and $K+1=O(\log(2\Lambda/\varepsilon))$ suffice.  The branch-selection rule and its robustness guarantee are given in Sec.~\ref{subsec:rpe-main} and Supplementary Sec.~\ref{app:phase-estimation}.  Allocating the failure probability across the $12N-9=O(N)$ coefficients and these scales gives a repetition count
\begin{equation}
    R_k=\left\lceil C\left(\log\!\frac{2(12N-9)}{\delta}+K-k\right)\right\rceil
    \label{eq:proof-sketch-repetitions}
\end{equation}
at scale $k$, for a sufficiently large numerical constant $C$.

One repetition at scale $k$ uses $O(N)$ robust-SWAP layers for initialization, encoding time $t_k$, and $O(N)$ additional robust-SWAP layers for alignment and readout.  Counting only native-Hamiltonian evolution as in Eq.~\eqref{eq:total-time-resource}, its interrogation time is $t_k+O(NJ_-^{-1}\log N)$. The $24$ Hamiltonian families and the two measurement bases contribute only a constant factor. Consequently,
\begin{equation}
    T_{\mathrm{tot}}=O\!\left(\sum_{k=0}^{K}R_k\left[t_k+\frac{N}{J_-}\log N\right]\right).
    \label{eq:proof-sketch-time-sum}
\end{equation}
Using $t_k=2^kt_0$, $t_K=\Theta(1/\varepsilon)$, and Eq.~\eqref{eq:proof-sketch-repetitions}, the evolution-time contribution satisfies $\sum_kR_kt_k=O(\varepsilon^{-1}\log(N/\delta))$, while $\sum_kR_k=O((K+1)\log(N/\delta)+(K+1)^2)$. Substituting $K+1=O(\log(2\Lambda/\varepsilon))$ into Eq.~\eqref{eq:proof-sketch-time-sum} gives Eq.~\eqref{eq:robust-learning-time}. A union bound over all coefficients, scales, and measurement bases gives the success probability in Eq.~\eqref{eq:robust-learning-guarantee}.

\end{proof}

\subsection{Global control}\label{subsec:global-control-main}

The protocol in Box~\ref{box:parallel} was stated using individual single-qubit control. In some experimental settings, however, it is easier to address a fixed number of periodic sublattices, applying the same gate simultaneously to all qubits in one chosen sublattice. For example, the zigzag dark-spin geometry proposed in Ref.~\cite{YaoEtAl2012} motivates arranging spins in a repeating transverse pattern, so that a magnetic-field gradient perpendicular to the chain gives different sublattices distinct resonance frequencies and permits frequency-selective global pulses. For neutral atoms, a commensurate optical superlattice could similarly distinguish sites within each repeating unit cell, while spatially patterned optical control provides another possible route~\cite{GorshkovEtAl2008}. We present the construction for three independently addressable sublattices; the same approach extends to any fixed control period $q\geq3$, including period four, with only constant-factor changes in the learning complexity.

For $r\in\{0,1,2\}$, define the sublattice $V_r^{(3)}:=\{j: j\equiv r\pmod 3\}$, and assume that any single-qubit gate $U$ can be applied simultaneously to all qubits in one chosen sublattice:
\begin{equation}
    \mathsf U_r:=\bigotimes_{\substack{j\in[N]\\j\equiv r\ ({\rm mod}\ 3)}}U_j.
    \label{eq:period-three-global-control-main}
\end{equation}
Direct preparation, measurement, and reset remain restricted to the bright qubit at site $1$; the dark register may initially be in an arbitrary state. We also assume that the designated interaction on every reference bond has the common Pauli form $g_iZ_iZ_{i+1}$, with $J_-\leq|g_i|\leq J_+$. This common-axis condition allows the same global control sequence to implement the robust SWAP construction on different bonds.

\paragraph{Select and reshape.} The choice of the Ramsey scale and measurement basis in the Select step of Box~\ref{box:parallel} is classical and requires no additional qubit control. As a first step, twirl the sublattice $r=0$ by applying the four global Pauli operations $\mathsf I_0,\mathsf X_0,\mathsf Y_0,$ and $\mathsf Z_0$ with equal weight. Assuming for simplicity that $N$ is divisible by three, the resulting Hamiltonian is
\begin{equation}
\begin{split}
    H_{\mathrm{eff}}^{(0)}&=\frac{1}{4}\sum_{P\in\{I,X,Y,Z\}}\mathsf P_0H\mathsf P_0^\dagger\\
    &=\sum_{q=0}^{N/3-1}\sum_{\alpha\in\{X,Y,Z\}}\left(h_{3q+1}^\alpha\sigma_{3q+1}^\alpha+h_{3q+2}^\alpha\sigma_{3q+2}^\alpha\right)+\sum_{q=0}^{N/3-1}\sum_{\alpha,\beta\in\{X,Y,Z\}}J_{3q+1,3q+2}^{\alpha\beta}\sigma_{3q+1}^\alpha\sigma_{3q+2}^\beta.
    \label{eq:three-patch-reshaping}
\end{split}
\end{equation}
Every one- or two-body nearest-neighbor term intersects the sublattice $r=0$ on at most one qubit. The Pauli twirl removes any term acting nontrivially on that qubit and preserves every term supported entirely on the remaining sites. Consequently, Eq.~\eqref{eq:three-patch-reshaping} decomposes the chain into the disjoint two-qubit patches $(3q+1,3q+2)$. Choosing $r=1$ or $r=2$ translates the same patch construction along the chain. 

The first twirl retains all Pauli terms supported within each patch. To isolate one common Pauli component across the patches, we apply the same additional twirl to every patch. Fix a nonidentity operator $Q\in\{I,X,Y,Z\}^{\otimes 2}\setminus\{I\otimes I\}$, let
\begin{equation}
    \mathcal C(Q):=\left\{P\otimes R\in\{I,X,Y,Z\}^{\otimes 2}:[P\otimes R,Q]=0\right\}
    \label{eq:patch-Pauli-centralizer}
\end{equation}
be its Pauli centralizer, with irrelevant global phases omitted. Using the period-three controls on the two sites of every patch, define the further reshaped Hamiltonian
\begin{equation}
    H_Q:=\frac{1}{|\mathcal C(Q)|}\sum_{P\otimes R\in\mathcal C(Q)}\mathsf P_1\mathsf R_2H_{\mathrm{eff}}^{(0)}\mathsf R_2^\dagger\mathsf P_1^\dagger.
    \label{eq:global-centralizer-reshaping}
\end{equation}
A Pauli operator survives this average only if it commutes with every element of $\mathcal C(Q)$. For a nonidentity two-qubit Pauli $Q$, the only such Pauli operators are $I\otimes I$ and $Q$. The effective Hamiltonian therefore has the form
\begin{equation}
    H_Q=\sum_{q=0}^{N/3-1}\lambda_{q,Q}Q_{3q+1,3q+2},
    \label{eq:globally-isolated-Pauli-family}
\end{equation}
where $Q_{3q+1,3q+2}$ denotes $Q$ acting on patch $(3q+1,3q+2)$ and $\lambda_{q,Q}$ is the corresponding coefficient $h_{3q+1}^\alpha$, $h_{3q+2}^\beta$, or $J_{3q+1,3q+2}^{\alpha\beta}$, according to the support of $Q$. Thus the same global pulse distribution isolates the same Pauli component on every disjoint patch while retaining its spatially varying coefficients.

\paragraph{Initialize.} Period-three control suffices to implement robust SWAPs simultaneously on any one of the three edge matchings. Define
\begin{equation}
    E_r^{(3)}:=\{(i,i+1):i\equiv r\pmod 3\},
    \qquad
    \widetilde S_r^{(3)}
    :=\prod_{(i,i+1)\in E_r^{(3)}}
    \widetilde{\operatorname{SWAP}}_{i,i+1},
    \quad r\in\{0,1,2\}.
    \label{eq:period-three-swap-layers-main}
\end{equation}
The reference interactions on each matching can be isolated using sublattice controls, and the same QSP sequence implements every SWAP in the layer despite their different unknown coupling strengths. To initialize the register, repeatedly apply $\widetilde S_0^{(3)}$, then $\widetilde S_2^{(3)}$, and then $\widetilde S_1^{(3)}$, resetting the bright qubit before the first cycle and after each cycle. With exact SWAPs, this sweep cycles through all sites, bringing every original subsystem to the bright qubit for reset. Robust SWAPs therefore prepare an approximation to the all-zero state without individual addressing or direct initialization of the dark qubits.

\paragraph{Encode.} Because the same Pauli operator $Q$ is selected on every patch, the Clifford gates that map its nonidentity factors to $Z$ are also the same on each corresponding sublattice and can be applied using Eq.~\eqref{eq:period-three-global-control-main}. The record qubits also occupy a single sublattice, so they can be prepared together by a sublattice Hadamard operation, denoted by $\mathsf{Had}_r$. After mapping the selected Pauli factors to $Z$, evolution under the reshaped Hamiltonian encodes the coefficients into separate phase records, just as in the individually controlled protocol.

\paragraph{Align and drain.} The layers in Eq.~\eqref{eq:period-three-swap-layers-main} also suffice to align and read out the records. After alignment, the records occupy sites $1,4,7,\ldots$. Measure and reset the bright qubit, then apply $\widetilde S_0^{(3)}$, followed by $\widetilde S_2^{(3)}$ and $\widetilde S_1^{(3)}$. In the ideal circuit, the next record follows $4\to3\to2\to1$, while each later record advances by three sites. Repeating this cycle reads out all records through the bright qubit using only sublattice control.

\paragraph{Estimate.} The Estimate step is unchanged. Measurements in the $X$ and $Y$ bases are performed only on the bright qubit, and the repetitions over the selected Hamiltonians and Ramsey scales are classical choices. The resulting two quadratures are processed by the same robust phase estimator as in Box~\ref{box:parallel}. The global-control construction changes the number of reshaping configurations and routing layers only by constant factors, so it leaves the asymptotic learning complexity unchanged. Supplementary Sec.~\ref{appendix:global_control} gives the pulse averages, state-preparation choices, and routing schedules in detail.

\section{Extension to bounded-degree graphs}
\label{sec:general-graphs}

We now extend the learning protocol to bounded-degree interaction graphs. Graph colorings replace the odd--even coefficient groups, and a spanning tree connects the dark qubits to the bright qubit. We first describe how records are routed through this tree, then combine routing with initialization and phase encoding.

Let $G=(V,E)$ be a known connected simple graph with $|V|=N$. Fix an ordering of the endpoints of each edge to specify the Pauli indices. The Hamiltonian is 
\begin{equation}
 H(\boldsymbol\lambda)
 =\sum_{i\in V}\sum_{\alpha\in\{X,Y,Z\}}
 h_i^\alpha\sigma_i^\alpha
 +\sum_{e=(i,j)\in E}\sum_{\alpha,\beta\in\{X,Y,Z\}}
 J_{ij}^{\alpha\beta}\sigma_i^\alpha\sigma_j^\beta .
 \label{eq:graph-Hamiltonian}
\end{equation}
There are $3N+9|E|$ Hamiltonian coefficients in total.  The coefficient promise in Eq.~\eqref{eq:coefficient-bound} is extended to
\begin{equation}
 \begin{aligned}
 |h_i^\alpha|&\leq\Lambda
 &&\text{for all }i\in V,\ \alpha\in\{X,Y,Z\},\\
 |J_{ij}^{\alpha\beta}|&\leq\Lambda
 &&\text{for all }(i,j)\in E,\
 \alpha,\beta\in\{X,Y,Z\}.
 \end{aligned}
 \label{eq:graph-coefficient-bound}
\end{equation}
We retain the access model of Sec.~\ref{subsec:access-model}: single-qubit control is available at every vertex, while direct preparation, measurement, and reset are restricted to the bright qubit $v_0$. The dark register may initially be in an arbitrary state. We assume that the maximum degree $\Delta_G$ is bounded independently of $N$, so the number of unknown coefficients is $O(N)$.

We assume that the reference backbone is a known spanning tree $T=(V,E_T)$ rooted at $v_0$. On every tree edge $e=(i,j)\in E_T$, a specified Pauli component has known axes $a_e,b_e\in\{X,Y,Z\}$ and an unknown coefficient in the interval $[J_-,J_+]$ in magnitude:
\begin{equation}
 P_e^{\rm ref}=\sigma_i^{a_e}\sigma_j^{b_e},
 \qquad
 g_e:=J_{ij}^{a_eb_e},
 \qquad
 J_-\leq |g_e|\leq J_+,\qquad e\in E_T.
 \label{eq:graph-reference-backbone}
\end{equation}
For each non-root vertex $v$, let $\operatorname{par}(v)$ denote its parent in the tree $T$ and define its level recursively by $\operatorname{lev}(v_0)=0$ and $\operatorname{lev}(v)=\operatorname{lev}(\operatorname{par}(v))+1$. The height of the rooted reference tree is
\begin{equation}
 h_T:=\max_{v\in V}\operatorname{lev}(v).
 \label{eq:reference-tree-height}
\end{equation}
We have $h_T\leq N-1$, with equality for a chain rooted at an endpoint. Transport and initialization use only the reference edges in $T$. 

\paragraph{Coefficient groups.} Vertex and edge colorings replace the odd--even partition used on the chain. Partition $V$ into proper vertex-color classes $V_1,\ldots,V_{C_V}$ and $E$ into proper edge-color classes $E_1,\ldots,E_{C_E}$. Thus no two vertices in one $V_c$ are adjacent, and no two edges in one $E_c$ share a common endpoint. A greedy vertex coloring uses $C_V\leq\Delta_G+1$ colors. A greedy edge coloring uses $C_E\leq2\Delta_G-1$ colors. Thus they are both bounded by constants independent of $N$.

Using the same notation as in Eqs.~\eqref{eq:one-body} and~\eqref{eq:two-body}, define the one-body and two-body Hamiltonian families as
\begin{equation}
\begin{split}
    \mathfrak{G}_{1}&=\left\{H_c^\alpha=\sum_{i\in V_c}h_i^\alpha\sigma_i^\alpha:c\in \{1,\ldots,C_{V}\},\alpha\in\{X,Y,Z\}\right\},\\
    \mathfrak{G}_{2}&=\left\{H_c^{\alpha\beta}=
 \sum_{(i,j)\in E_c}J_{ij}^{\alpha\beta}
 \sigma_i^\alpha\sigma_j^\beta: c\in\{1,\ldots,C_{E}\},\alpha,\beta\in\{X,Y,Z\}\right\}.
 \label{eq:graph-encoding-groups}
\end{split}
\end{equation}
Then
\begin{equation}
 H=\sum_{c=1}^{C_V}\sum_\alpha H_c^\alpha
 +\sum_{c=1}^{C_E}\sum_{\alpha,\beta}H_c^{\alpha\beta}.
 \label{eq:graph-group-decomposition}
\end{equation}
Define $\mathfrak{G}=\mathfrak{G}_1\cup\mathfrak{G}_2$. The total number of Hamiltonians in $\mathfrak{G}$ is
\begin{equation}
 3C_V+9C_E=O(\Delta_G),
 \label{eq:number-graph-groups}
\end{equation}
which is constant for bounded degree. The terms within each selected Hamiltonian have disjoint supports and can be isolated simultaneously by Hamiltonian reshaping.

For a one-body Hamiltonian, every selected vertex is a record site. For a two-body Hamiltonian, choose one endpoint of each selected edge as the record and the other as the anchor. Write $A=\sum_{j\in I_A}\lambda_jP_j$ for a selected Hamiltonian, let $r_j$ be its record sites, and define $R_A:=\{r_j:j\in I_A\}$.

\paragraph{Tree routing.} The record sites produced by a graph group need not lie on a regularly spaced sublattice, so the fixed chain cycle $D=S_oS_e$ is replaced by the following schedule.  It is a special case of matching routing with consumption~\cite{PantziouRobertsSymvonis1996}. We state the construction explicitly because here each vertex contains only one qubit and consumption is implemented by measurement and reset.

\begin{lemma}[Rooted-tree drain]
\label{lem:tree-drain}
Let $T$ be rooted at the bright vertex $v_0$. Suppose the initial state is the tensor product of single-qubit record states $\rho_1,\ldots,\rho_\nu$ on distinct non-root vertices $r_1,\ldots,r_\nu$, with $\nu\geq1$, and the state on every other vertex is $\ket0\!\bra0$. Define the largest level occupied by a record as
\begin{equation}
 h_\star:=\max_{1\leq a\leq\nu}\operatorname{lev}(r_a).
\end{equation}
Then a predetermined sequence of matching SWAP layers delivers and removes all records at $v_0$ within
\begin{equation}
 L_{\rm drain}\leq h_\star+2\nu-2
 \label{eq:tree-drain-depth}
\end{equation}
layers. A record initially at $v_0$ can be measured before the schedule begins and is not included in $\nu$.
\end{lemma}

\begin{proof}
Order the record sites $r_1,\ldots,r_\nu$ so that an ancestor precedes each of its descendants; restricting a preorder traversal of $T$ to the record sites gives such an order. Let $h_a:=\operatorname{lev}(r_a)$ and write the unique path from $r_a$ to the root as
\begin{equation}
 r_a=x_{a,h_a},x_{a,h_a-1},\ldots,x_{a,0}=v_0,
\end{equation}
where $\operatorname{lev}(x_{a,j})=j$. For $j=h_a,h_a-1,\ldots,1$, schedule
\begin{equation}
 \operatorname{SWAP}_{x_{a,j},x_{a,j-1}}
 \quad\text{in layer}\quad
 \tau(a,j):=h_\star+2a-j-1.
 \label{eq:tree-drain-schedule}
\end{equation}
The layers for a fixed record are consecutive because $\tau(a,j-1)=\tau(a,j)+1$, and its root-edge SWAP occurs in layer $h_\star+2a-2$.

We next verify that every scheduled layer is a matching. If two tree edges share a vertex, the levels $j$ and $j'$ of their child endpoints differ by at most one. Equality of their scheduled layers would imply
\begin{equation}
 2(a-a')=j-j'.
\end{equation}
The left side is even, whereas $j-j'\in\{-1,0,1\}$. Hence $j=j'$ and then $a=a'$. A root path contains only one edge whose child has level $j$, so the two scheduled operations coincide rather than conflict.

Finally, the schedule never exchanges two unread records. Once a route has started, its record occupies a strict internal vertex at level $h$ only between the consecutive entry and departure layers $\tau(a,h+1)$ and $\tau(a,h)$. If two routes use that vertex only in transit, their entry layers differ by $2|a-a'|\geq2$, so these one-layer visits do not overlap. The remaining case is that the vertex is itself the source of an ancestor record. If $r_a$ is an ancestor of $r_{a'}$, then $a<a'$, and
\begin{equation}
 \tau(a',h_a+1)-\tau(a,h_a)=2(a'-a)-1\geq1;
\end{equation}
the ancestor has therefore departed before the descendant enters its source. These cases cover every common non-root vertex of two root paths. At the root, record $a$ arrives in layer $h_\star+2a-2$, so successive arrivals are separated by two layers, and the root can be measured and reset between them. Induction over the scheduled moves now shows that the destination of every SWAP is a blank qubit. Exact SWAPs merely permute the labelled product-state factors, so the schedule requires no measurement of record occupancy. The last record arrives in layer $h_\star+2\nu-2$, which proves Eq.~\eqref{eq:tree-drain-depth}; Supplementary Sec.~\ref{app:tree-routing} gives the detailed occupancy induction.
\end{proof}

This bound captures both limitations of a single-probe drain. A record initially at level $h_\star$ must cross $h_\star$ tree edges, while the root can receive at most one new record in each matching layer. These observations give the lower bound $\Omega(\max\{h_\star,\nu\})=\Omega(h_\star+\nu)$, matching Lemma~\ref{lem:tree-drain} up to a constant factor.

\paragraph{Initialization and encoding.} Every matching of reference-tree edges can be implemented using robust SWAPs. Hamiltonian reshaping isolates the designated reference components, local Clifford controls map their known axes to $ZZ$, and Theorem~\ref{thm:robust_swap} supplies the common QSP sequence. We use these operations both to initialize the register and to implement the routing schedule above.

Initialization uses a fixed sweep through the tree edges. Restrict the edge-color classes to $T$, setting $M_c:=E_c\cap E_T$, and define
\begin{equation}
    S_c^T:=\prod_{e\in M_c}\operatorname{SWAP}_e,
    \qquad D_T:=S_{C_E}^T\cdots S_1^T.
    \label{eq:tree-initialization-sweep}
\end{equation}
Each tree edge appears exactly once in this sweep, and $D_T$ is therefore a single permutation cycle through all $N$ vertices~\cite{Denes1959}. Reset the bright qubit, then apply the robust versions of $S_1^T,\ldots,S_{C_E}^T$ and reset it again, repeating this sweep-and-reset cycle $N-1$ times. In the ideal circuit, every original subsystem is discarded and the register is left in $\ket0^{\otimes N}$, even for an initially entangled input. The robust circuit gives approximate initialization through the same mechanism as Box~\ref{box:initialization}.

As on the chain, conjugating the reshaped evolution by local Clifford gates maps the selected terms to $Z$ or $ZZ$. After initialization, apply Hadamard gates $\mathrm{Had}_{r_j}$ to the record sites and evolve under this effective Hamiltonian for time $t$. For the ideal initialized state, this produces
\begin{equation}
    \ket{\Psi_A(t)}
    =\bigotimes_{j\in I_A}
      \left(e^{-i\lambda_jtZ}\ket+\right)_{r_j}
      \otimes\ket0_{V\setminus R_A},
    \label{eq:graph-ideal-record-state}
\end{equation}
with the same Ramsey signals as in Eq.~\eqref{eq:parallel-family-signals}. The implemented encoded state inherits the initialization error, which is bounded together with the routing errors. The tree schedule replaces the chain-specific Align and Drain steps; the choices of Hamiltonian family, Ramsey time, and measurement basis, and the final estimation procedure, remain as in Box~\ref{box:parallel}.

\paragraph{Learning guarantee.} Let $\nu_A$ be the number of non-root records in one Hamiltonian $A\in\mathfrak{G}$. At a fixed evolution time, measurement basis, and statistical repetition, applying Lemma~\ref{lem:tree-drain} to every nonempty group gives the SWAP-layer depth of one complete coefficient sweep,
\begin{align}
 L_{\rm route}
 &\leq \sum_{A\in\mathfrak{G}}
 \bigl[h_T+2\nu_A\bigr] \notag\\
 &\leq
 \bigl(3C_V+9C_E\bigr)h_T
 +6N+18|E|.
 \label{eq:graph-routing-depth}
\end{align}
Since $h_T\leq N-1$ and the handshaking identity gives $2|E|\leq N\Delta_G$, Eq.~\eqref{eq:graph-routing-depth} implies
\begin{equation}
\label{eq:linear-graph-routing}
    L_{\rm route}=O(N\Delta_G).
\end{equation}
Initialization uses $C_E(N-1)$ matching layers before each encoding experiment. Since both $C_E$ and the number of selected Hamiltonians are constant for bounded degree, initialization and readout together still require $O(N)$ layers per complete sweep.

Applying the error analysis and phase-estimation argument of Theorem~\ref{thm:robust-parallel-learning}, with initialization included in every repetition, gives the following guarantee.

\begin{theorem}[Robust learning on bounded-degree graphs]
\label{thm:robust-graph-learning}
Let $G$ be a known connected simple interaction graph whose maximum degree $\Delta_G$ is bounded independently of $N$. Suppose Eq.~\eqref{eq:graph-coefficient-bound} holds, and a known spanning tree $T\subseteq G$ satisfies the reference-backbone condition in Eq.~\eqref{eq:graph-reference-backbone}. Under the individual-control model, for any $0<\varepsilon<\Lambda$ and $0<\delta<1$, the parallel protocol learns all $3N+9|E|$ coefficients to $\ell_\infty$ error $\varepsilon$ with probability at least $1-\delta$. It suffices to implement each tree-edge SWAP channel with diamond-norm error $\eta=O(N^{-2})$, and the total native-Hamiltonian evolution time is
\begin{equation}
\begin{split}
 T_{\rm tot}=O\!\left[
 \frac{1}{\varepsilon}\log\!\left(\frac{N\Delta_{G}}{\delta}\right)
 +\frac{N\Delta_{G}}{J_-}\log N\left(
 \log\!\left(\frac{N\Delta_{G}}{\delta}\right)
 +\log\!\left(\frac{2\Lambda}{\varepsilon}\right)
 \right)
 \log\!\left(\frac{2\Lambda}{\varepsilon}\right)
 \right].
 \label{eq:robust-graph-learning-time}
\end{split}
\end{equation}
In particular, the precision dependence remains Heisenberg limited, and for fixed $\varepsilon$, $\delta$, $\Lambda$, $J_\pm$, and $\Delta_G$, the total time is $\widetilde O(N)$.
\end{theorem}

\begin{proof}[Proof sketch]
The colorings make the number of experimental groups depend only on $\Delta_G$, while the ideal factorization in Eq.~\eqref{eq:graph-ideal-record-state} gives the same Ramsey signals as on the chain. Lemma~\ref{lem:tree-drain} and Eq.~\eqref{eq:graph-routing-depth} give $O(N)$ readout layers per complete sweep for bounded degree. The fixed tree sweeps also give $O(N)$ initialization layers per complete sweep. The triangle inequality, including both initialization and readout errors, keeps every Ramsey marginal within a constant bias when $\eta=O(N^{-2})$. The multiscale estimator and confidence allocation are then identical to the proof of Theorem~\ref{thm:robust-parallel-learning}. Substituting the layer time $O(J_-^{-1}\log N)$ and the combined initialization and readout depth into the sum over phase scales and repetitions gives Eq.~\eqref{eq:robust-graph-learning-time}. Supplementary Sec.~\ref{app:tree-routing} supplies the detailed routing, error, and time bounds.
\end{proof}

For bounded degree, this matches the chain complexity. The graph theorem uses individual control to select a different physical matching in each layer.  The three-sublattice construction of Sec.~\ref{subsec:global-control-main} establishes a weaker-control result for the chain, but does not by itself implement the irregular tree schedule above.  Extending Eq.~\eqref{eq:linear-graph-routing} under strictly global control therefore requires additional symmetry or addressability assumptions on the graph.


\section{Discussion and outlook}
\label{sec:discussion}
In this work, we have established an efficient protocol for learning an extensive local Hamiltonian when measurements and resets are restricted to a single qubit. The protocol addresses two technical challenges: constructing SWAP gates from unknown interactions and organizing learning efficiently. Quantum signal processing resolves the first challenge by synthesizing parameter-independent robust SWAPs. Hamiltonian reshaping, parallel phase encoding, and multiscale phase estimation then allow many coefficients to be learned in parallel. The resulting protocol retains Heisenberg-limited precision scaling and has
$\widetilde O(N)$ total evolution time at fixed accuracy and confidence.  The scaling also holds for chains with periodic global control and for general bounded-degree graphs. On chains, it matches the fundamental precision and propagation limits up to logarithmic factors, with the physical parameters held fixed.

The various interaction geometries of the many-body Hamiltonian provide a natural direction for further study. For a general bounded-degree interaction graph, we choose a spanning tree rooted at the measured qubit and use it to route information from across the system to the measurement region, while couplings outside the tree are isolated and learned through Hamiltonian reshaping. This construction retains near-linear learning complexity, but our current bounds depend conservatively on properties of the chosen tree, including its depth and the congestion of transport paths near the root. An important open question is therefore how the interaction geometry determines the optimal learning complexity. Even for a fixed graph, the choice of measurement location and transport tree can substantially alter the distance and congestion of information routing. Allowing multiple measured qubits or a finite readout region introduces further tradeoffs among probe placement, graph distance, routing congestion, and total learning time. Characterizing these tradeoffs, and extending efficient routing on irregular graphs to settings with global or symmetry-constrained control, would provide a more complete understanding of the role of geometry in local-probe Hamiltonian learning.

Our analysis counts native Hamiltonian evolution time and treats single-qubit operations, measurement,
reset, and the short-time limit used in Hamiltonian reshaping as ideal. Therefore, an immediate open problem is to understand how these guarantees change under realistic control and noise constraints.  It would be valuable to include finite pulse durations, bounded pulse rates, and decoherence during both phase accumulation and transport,
and to determine when the Heisenberg scaling survives.  Platform-specific noise models and numerical pulse optimization may also substantially reduce the conservative overhead associated with worst-case robust-SWAP errors.

Conceptually, our results identify local-probe Hamiltonian learning as a distinct learning paradigm rather than merely a constrained version of
distributed measurement.  Measurement locality becomes an independent resource, and coherent control converts spatially remote parameters into signals available at a fixed interface.  The fact that one measured qubit can recover an extensive set of Hamiltonian coefficients with nearly optimal
scaling suggests that local probes can serve not only as sensors of nearby degrees of freedom, but as scalable interfaces for learning, verifying, and controlling complex quantum systems.

\section*{Acknowledgement}
We thank Yu Tong for insightful discussions. Z.X acknowledges support from QuICS Lanczos Graduate Fellowship. S.L acknowledges support from a QuICS seed grant.
Y.-X.W.~acknowledges support from a QuICS Hartree Postdoctoral Fellowship. 
A.V.G.~and Z.-Y.W. acknowledge support from the U.S.~Department of Energy, Office of Science, Accelerated Research in Quantum Computing, Fundamental Algorithmic Research toward Quantum Utility (FAR-Qu). A.V.G.~and Z.-Y.W.~were also supported in part by the DoE ASCR Quantum Testbed Pathfinder program (award No.~DE-SC0024220), ONR MURI, NSF QLCI (awards No.~OMA-2120757 and No.~OSI-2553574), NSF STAQ program, AFOSR MURI, ARL (W911NF-24-2-0107), and NQVL:QSTD:Design:FTL. A.V.G.~and Z.-Y.W.~also acknowledge support from the U.S.~Department of Energy, Office of Science, National Quantum Information Science Research Centers, Quantum Systems Accelerator (award No.~DE-SCL0000121). 

ChatGPT was used to assist with language editing and the refinement of some proofs. The research ideas, methods, and results were developed by the authors. All AI-assisted mathematical arguments were carefully checked and verified by the authors, who take full responsibility for the correctness and content of the manuscript.

\bibliography{main}
\clearpage

\onecolumngrid
\begin{center}
\textbf{\large Supplementary Material: }
\end{center}

\setcounter{section}{0}
\setcounter{equation}{0}
\setcounter{figure}{0}
\setcounter{table}{0}
\setcounter{page}{1}
\renewcommand{\theequation}{S\arabic{equation}}
\renewcommand{\thefigure}{S\arabic{figure}}
\renewcommand{\thetable}{S\arabic{table}}

This Supplementary Material provides the detailed derivations and proofs supporting the main text. Section~\ref{app:phase-estimation} derives the robust phase-estimation procedure and its complexity. Section~\ref{appendix:robust_swap} constructs robust SWAP gates using quantum signal processing. Section~\ref{app:marginal-lemma} proves the marginal-error bound, and Sec.~\ref{app:robust-learning-proof} establishes the error and time guarantees for the learning protocol. Section~\ref{appendix:global_control} describes control through three sublattices. Finally, Sec.~\ref{app:tree-routing} gives the routing and initialization proofs for general graphs.

\section{Robust phase estimation}
\label{app:phase-estimation}

This section derives the estimation rule and repetition schedule used in Sec.~\ref{subsec:rpe-main} and Box~\ref{box:parallel}. We use the same unknown coefficient $\lambda$, prior bound $|\lambda|\leq\Lambda$, and Ramsey phase $2\lambda t$ as in the main text. The argument has three parts: repeated measurements determine each phase to constant accuracy despite a bounded systematic bias; measurements at successively doubled times resolve the phase ambiguity; and an appropriate allocation of repetitions gives the stated total evolution time.

\subsection{From measurement probabilities to a phase estimate}

For a qubit prepared in $\ket+$, evolution under $\lambda Z$ produces
\begin{equation}
    \ket{\psi(t)}
    =\frac{e^{-i\lambda t}\ket0+e^{i\lambda t}\ket1}{\sqrt2}.
\end{equation}
Its off-diagonal density-matrix element is $\rho_{01}=e^{-2i\lambda t}/2$. Consequently, $\langle X\rangle_t=2\operatorname{Re}\rho_{01}=\cos(2\lambda t)$ and $\langle Y\rangle_t=-2\operatorname{Im}\rho_{01}=\sin(2\lambda t)$. At time $t_k$, write
\begin{equation}
    \varphi_k:=2\lambda t_k,\qquad
    c_k:=\cos\varphi_k,\qquad s_k:=\sin\varphi_k.
    \label{eq:supp-phase-quadratures}
\end{equation}
The ideal $+1$ outcome probabilities are $p_{X,k}=(1+c_k)/2$ and $p_{Y,k}=(1+s_k)/2$. As in Eq.~\eqref{eq:rpe-biased-probabilities}, the actual probabilities may differ:
\begin{equation}
    \widetilde p_{B,k}=p_{B,k}+b_{B,k},
    \qquad |b_{B,k}|\leq b,\qquad B\in\{X,Y\}.
\end{equation}
The bias may depend on the basis, scale, and coefficient being estimated. We require only a uniform bound at all sampled times, including the longest one.

Repeat the complete experiment independently $R_k$ times for each basis. Let $n_{B,k}$ count the $+1$ outcomes and define $\widehat p_{B,k}:=n_{B,k}/R_k$. These empirical probabilities estimate $\widetilde p_{B,k}$, so repetition reduces sampling error but does not remove the systematic bias. For a chosen sampling tolerance $a>0$, define the event
\begin{equation}
    \mathcal E_k:=
    \left\{|\widehat p_{X,k}-\widetilde p_{X,k}|\leq a,\
    |\widehat p_{Y,k}-\widetilde p_{Y,k}|\leq a\right\}.
\end{equation}
On this event, the triangle inequality gives $|\widehat p_{B,k}-p_{B,k}|\leq a+b$. The estimated quadratures $\widehat c_k:=2\widehat p_{X,k}-1$ and $\widehat s_k:=2\widehat p_{Y,k}-1$ therefore satisfy
\begin{equation}
    |\widehat c_k-c_k|,\ |\widehat s_k-s_k|
    \leq2(a+b)\leq\frac18
    \quad\text{when}\quad a=\frac1{32},\quad b\leq\frac1{32}.
    \label{eq:rpe-quadrature-error}
\end{equation}
These are the constants used in the later learning proofs. They are sufficient bounds, rather than optimal noise thresholds.

To convert this quadrature bound into a phase bound, introduce $z_k:=c_k+is_k=e^{i\varphi_k}$ and $\widehat z_k:=\widehat c_k+i\widehat s_k$. Equation~\eqref{eq:rpe-quadrature-error} implies
\begin{equation}
    |\widehat z_k-z_k|
    \leq\sqrt{(\widehat c_k-c_k)^2+(\widehat s_k-s_k)^2}
    \leq\frac{\sqrt2}{8}=:r<1.
\end{equation}
In particular, $|\widehat z_k|\geq1-r>0$, so $\widehat\varphi_k:=\operatorname{Arg}(\widehat z_k)\in(-\pi,\pi]$ is well defined on $\mathcal E_k$. If $\widehat z_k=0$ on a failed sampling event, the estimator may assign any phase, say zero.

Define circular phase distance by $d_{\rm circ}(\theta,\phi):=\min_{\ell\in\mathbb Z}|\theta-\phi+2\pi\ell|$. Let $\vartheta_k\in[-\pi,\pi]$ be the signed phase difference and write $e^{-i\varphi_k}\widehat z_k=\rho e^{i\vartheta_k}$ with $\rho>0$. Since $|e^{-i\varphi_k}\widehat z_k-1|\leq r$, its real part is at least $1-r>0$, and hence $|\vartheta_k|<\pi/2$. Moreover,
\begin{equation}
    r^2\geq|\rho e^{i\vartheta_k}-1|^2
    =(\rho-\cos\vartheta_k)^2+\sin^2\vartheta_k
    \geq\sin^2\vartheta_k.
\end{equation}
It follows that
\begin{equation}
    d_{\rm circ}(\widehat\varphi_k,\varphi_k)
    =|\vartheta_k|
    \leq\arcsin r
    =\arcsin\frac{\sqrt2}{8}
    <\frac{\pi}{6}.
    \label{eq:supp-rpe-phase-error}
\end{equation}
Thus we may use the fixed phase-error bound $\gamma=\pi/6$. More generally, the same calculation gives a bound $\arcsin[2\sqrt2(a+b)]$ whenever $2\sqrt2(a+b)<1$.

\subsection{Resolving the phase ambiguity}

Choose $t_0:=\pi/(4\Lambda)$ and $t_k:=2^kt_0$ for $k=0,\ldots,K$. We first prove a deterministic statement: if every inferred phase obeys $d_{\rm circ}(\widehat\varphi_k,2\lambda t_k)\leq\gamma$ for some $0<\gamma<\pi/3$, then the rule in Eq.~\eqref{eq:rpe-unwrapping-main} gives
\begin{equation}
    |\widehat\lambda_k-\lambda|
    \leq\frac{\gamma}{2t_k}
    \qquad (k=0,\ldots,K).
    \label{eq:supp-frequency-error}
\end{equation}

At the initial scale, $|2\lambda t_0|\leq\pi/2$. Because $\gamma<\pi/3$, the entire interval within $\gamma$ of this true phase lies inside $(-\pi,\pi)$. Its principal-value estimate therefore satisfies the ordinary, rather than only circular, bound $|\widehat\varphi_0-2\lambda t_0|\leq\gamma$. Defining $\widehat\lambda_0:=\widehat\varphi_0/(2t_0)$ proves Eq.~\eqref{eq:supp-frequency-error} for $k=0$.

For $k\geq1$, every coefficient compatible with the measured phase has the form $(\widehat\varphi_k+2\pi\ell)/(2t_k)$. Select the one nearest the preceding estimate:
\begin{equation}
    \widehat\lambda_k
    :=\underset{\lambda'\in\{(\widehat\varphi_k+2\pi\ell)/(2t_k):\,\ell\in\mathbb Z\}}
    {\operatorname{argmin}}|\lambda'-\widehat\lambda_{k-1}|.
    \label{eq:supp-phase-unwrapping}
\end{equation}
The circular-error bound guarantees a candidate $\lambda_k^\star$ with $|\lambda_k^\star-\lambda|\leq\gamma/(2t_k)$. Assume inductively that Eq.~\eqref{eq:supp-frequency-error} holds at scale $k-1$. Since $t_k=2t_{k-1}$,
\begin{equation}
    |\widehat\lambda_{k-1}-\lambda|
    \leq\frac{\gamma}{2t_{k-1}}
    =\frac{\gamma}{t_k},
    \qquad
    |\lambda_k^\star-\widehat\lambda_{k-1}|
    \leq\frac{3\gamma}{2t_k}.
\end{equation}
Distinct candidates are separated by at least $\pi/t_k$. Thus every other candidate $\lambda'\neq\lambda_k^\star$ satisfies
\begin{equation}
    |\lambda'-\widehat\lambda_{k-1}|
    \geq|\lambda'-\lambda_k^\star|-|\lambda_k^\star-\widehat\lambda_{k-1}|
    \geq\frac{\pi}{t_k}-\frac{3\gamma}{2t_k}
    >\frac{3\gamma}{2t_k},
\end{equation}
where the strict inequality is precisely $3\gamma<\pi$. The nearest candidate is therefore uniquely $\lambda_k^\star$, which proves the induction. On failed sampling events, any ties can be resolved by a fixed convention.

With $\gamma=\pi/6$, the schedule in Box~\ref{box:parallel} is
\begin{equation}
    K:=\min\left\{q\in\mathbb Z_{\geq0}:2^qt_0\geq\frac{\pi}{12\varepsilon}\right\}.
    \label{eq:supp-rpe-final-scale}
\end{equation}
Equation~\eqref{eq:supp-frequency-error} then gives $|\widehat\lambda_K-\lambda|\leq\pi/(12t_K)\leq\varepsilon$. For $0<\varepsilon<\Lambda$, minimality of $K$ implies
\begin{equation}
    \frac{\pi}{12\varepsilon}\leq t_K
    \leq\max\left\{t_0,\frac{\pi}{6\varepsilon}\right\}
    \leq\frac{\pi}{4\varepsilon}.
\end{equation}
The middle bound also covers $K=0$. Hence $t_K=\Theta(1/\varepsilon)$ and $K+1=O(1+\log(2\Lambda/\varepsilon))$. This is why a fixed phase error suffices: dividing it by $2t_K$ gives the desired coefficient accuracy.

\subsection{Repetitions and joint success probability}

We now choose $R_k$ so that the sampling events used above hold with high probability. For independent repetitions in basis $B$, Hoeffding's inequality for the Bernoulli indicators of the $+1$ outcome gives
\begin{equation}
    \Pr\!\left(|\widehat p_{B,k}-\widetilde p_{B,k}|>a\right)
    \leq2e^{-2R_ka^2}.
\end{equation}
A union bound over the two bases therefore yields $\Pr(\mathcal E_k^c)\leq4e^{-2R_ka^2}$. If $\delta_k$ is the allowed failure probability at scale $k$, it suffices to take
\begin{equation}
    R_k=\left\lceil\frac{1}{2a^2}\log\frac4{\delta_k}\right\rceil.
    \label{eq:supp-rpe-exact-repetitions}
\end{equation}
All logarithms in this section are natural logarithms. For a target failure probability $0<\delta\leq1/2$, allocate
\begin{equation}
    \delta_k:=\delta\,2^{k-K-1},\qquad
    \sum_{k=0}^K\delta_k
    =\delta(1-2^{-K-1})<\delta.
\end{equation}
Substitution into Eq.~\eqref{eq:supp-rpe-exact-repetitions} gives
\begin{equation}
    R_k=\left\lceil\frac{1}{2a^2}
    \left[\log\frac1\delta+(K-k+3)\log2\right]\right\rceil
    =O\!\left(\log\frac1\delta+K-k\right).
    \label{eq:supp-rpe-repetition}
\end{equation}
Here $a=1/32$ is fixed, and $\log(1/\delta)\geq\log2$ absorbs the additive constants and the ceiling. Equivalently, the schedule $R_k=\lceil C[\log(1/\delta)+K-k]\rceil$ in Eq.~\eqref{eq:rpe-repetitions-main} suffices for a sufficiently large universal $C$. The union bound gives $\Pr(\bigcap_{k=0}^K\mathcal E_k)>1-\delta$. On this event, the phase bound and the deterministic induction prove $|\widehat\lambda_K-\lambda|\leq\varepsilon$.

For simultaneous estimation of $L$ coefficients, assign each coefficient failure budget $\delta/L$ and each of its scales budget $\delta_{j,k}:=(\delta/L)2^{k-K-1}$, where $j=1,\ldots,L$. The corresponding repetition count is
\begin{equation}
    R_k=O\!\left(\log\frac{L}{\delta}+K-k\right).
    \label{eq:supp-rpe-many-coefficients}
\end{equation}
Summing the failure probabilities over $j$ and $k$ gives a joint failure probability below $\delta$. Outcomes for different coefficients within one experimental shot may be correlated: the union bound does not require independence between them. Concentration requires independent repetitions of the complete experiment for each fixed coefficient, scale, and basis. This distinction allows the argument to apply to the approximate-SWAP readout protocol.

\subsection{Evolution time and the cost of repeated readout}

Let $B:=\log(L/\delta)$, initially with $\delta\leq1/2$, and choose the repetition schedule above. There is a constant $C_0$ such that $R_k\leq C_0(B+K-k)$ for every $k$. The evolution time for the two quadratures of one selected Hamiltonian is $T_{\rm phase}:=2\sum_{k=0}^K R_kt_k$. Setting $j:=K-k$ and using $t_k=2^{-j}t_K$ gives
\begin{align}
    T_{\rm phase}
    &\leq2C_0t_K\sum_{j=0}^K2^{-j}(B+j)\notag\\
    &\leq2C_0t_K\left(B\sum_{j=0}^\infty2^{-j}
                      +\sum_{j=0}^\infty j2^{-j}\right)\notag\\
    &=4C_0t_K(B+1)
    =O\!\left(\frac1\varepsilon\log\frac{L}{\delta}\right).
    \label{eq:supp-rpe-evolution-time}
\end{align}
The two series equal $2$: the first follows from $\sum_{j=0}^\infty x^j=(1-x)^{-1}$ at $x=1/2$, and the second follows by differentiating this identity and multiplying by $x$, giving $\sum_{j=0}^\infty jx^j=x/(1-x)^2$. Since $B\geq\log2$, $B+1=O(B)$. Taking $L=1$ recovers Eq.~\eqref{eq:rpe-time-main}. In parallel learning, $L$ enters this bound through the joint confidence requirement; coefficients encoded in the same experiment share the evolution time. A sweep over several selected Hamiltonians must additionally sum their encoding times.

Readout also incurs a routing cost each time the experiment is repeated. Unlike signal evolution, this cost is not weighted by $t_k$, so the later proofs need the separate bound
\begin{align}
    \sum_{k=0}^KR_k
    &\leq C_0\sum_{k=0}^K(B+K-k)\notag\\
    &=C_0\left[(K+1)B+\frac{K(K+1)}2\right]\notag\\
    &=O\!\left((K+1)\log\frac{L}{\delta}+K(K+1)\right).
    \label{eq:supp-rpe-total-repetitions}
\end{align}
Thus, if each repetition of a selected Hamiltonian requires initialization and routing time $T_{\rm prep+route}$ in addition to its encoding time, its contribution for both bases is
\begin{equation}
    2\sum_{k=0}^KR_k(t_k+T_{\rm prep+route})
    =O\!\left[
    \frac1\varepsilon\log\frac{L}{\delta}
    +T_{\rm prep+route}\left((K+1)\log\frac{L}{\delta}+K(K+1)\right)
    \right].
\end{equation}
These two sums are used in the chain and graph complexity proofs. The phase-estimation argument itself requires only the uniform probability-bias bound; those proofs establish that sufficiently accurate robust SWAPs satisfy it.

\section{Proof for Theorem~\ref{thm:robust_swap}}
\label{appendix:robust_swap}

We first prove the robust entangling rotation in Lemma~\ref{lem:robust_entangler}.

\begin{proof}
    Choose the free-evolution time $\tau=\pi/[2(J_-+J_+)]$ and define the accumulated angle $\theta_J=\pi J/(J_-+J_+)$. Its magnitude lies in the interval 
    \begin{equation}
        |\theta_J| \in [\theta_0,\pi-\theta_0],
        \qquad
        \theta_0 = \frac{\pi J_-}{J_-+J_+}.
    \label{eq:theta_interval}
    \end{equation}
    Consequently, 
    \begin{equation}
        |\sin\theta_J| \geq \sin\theta_0 \equiv \Delta ,
        \qquad
        \operatorname{sgn}(\sin\theta_J) = \operatorname{sgn}(J).
    \label{eq:sin_gap}
    \end{equation}
    A bounded polynomial approximation to the sign function follows from Corollary~6 of Ref.~\cite{LowChuang2017Uniform}, with a rescaling to ensure the bound below. For every $0<\Delta<1$ and $0<\zeta<1$, there exists an odd real polynomial $p$ of degree
    \begin{equation}
        d=O\!\left(\frac{1}{\Delta}\log\frac{1}{\zeta}\right)
        \label{eq:sign_degree}
    \end{equation}
    such that
    \begin{equation}
        |p(s)|\leq1\quad(s\in[-1,1]),
        \qquad
        |p(s)-\operatorname{sgn}(s)|\leq\zeta
        \quad(\Delta\leq|s|\leq1).
        \label{eq:sign_polynomial}
    \end{equation}
    This polynomial can be implemented using the Equiangular quantum transformation with the primitive rotations. 

    Set $\zeta: = \eta^2/16$. Since $p$ is an odd real polynomial, we can write it as $p(s) = sq(s^2)$. For $x = \cos(\theta/2)$ and $y = \sin(\theta/2)$, define 
    \begin{align}
    A(x) &:=1-4(1-\cos\alpha)x^2(1-x^2) q\!\left(4x^2(1-x^2)\right)^2,
    \label{eq:robust_response_A}
    \\
    C(y) &:=-2\sin\alpha\,y\, q\!\left(4y^2(1-y^2)\right).
    \label{eq:robust_response_C}
\end{align}
Then $A$ is an even polynomial with degree at most $4d$ and $C$ is an odd polynomial with degree at most $2d$. Moreover, $A(1) = 1.$ Also, 
\begin{align}
    A(x)^2 + x^2 C(y)^2 & = 1-(1-\cos\alpha)^2u^2(1-u^2) \le 1, 
    \label{eq:robust_partial_tuple_bound}
\end{align}
as $u : = p(\sin\theta)$ with $|u| \le 1$.

Theorem~2(1) of Ref.~\cite{Low2016Methodology}, together with its constructive completion in Lemma~3 and the even-length extension, supplies polynomials $B$ and $D$ with the required parity. Lemma~\ref{lem:equiangular_response} then implements $(A,B,C,D)$ with an even-length sequence of $L=4d$ rotations. It remains to convert the polynomial approximation error to an operator-norm bound. For $J_-\leq|J|\leq J_+$, set $s:=\sin\theta_J$ and $r:=\operatorname{sgn}(s)=\operatorname{sgn}(J)$. From Eq.~\eqref{eq:sign_polynomial}, $|u-r|\leq\zeta$ and $1-u^2\leq2\zeta$. Hence
\begin{equation}
    A(x)=1-(1-\cos\alpha)u^2,
    \qquad
    xC(y)=-\sin\alpha\,u.
    \label{eq:robust_response_pair}
\end{equation}
while unitarity and Eq.~\eqref{eq:robust_partial_tuple_bound} give
\begin{equation}
    B(x)^2+x^2D(y)^2 =(1-\cos\alpha)^2u^2(1-u^2) \leq8\zeta.
\end{equation}

Therefore,
\begin{align}
    \left\|
        U_{\boldsymbol\phi}(\theta)
        -e^{-ir\alpha\overline X}
    \right\|^2
    & = 
    |A(x)-\cos\alpha|^2 +|xC(y)+r\sin\alpha|^2 +B(x)^2+x^2D(y)^2
    \notag\\
    &\leq17\zeta^2+8\zeta
    <\eta^2.
    \label{eq:robust_operator_error}
\end{align}
Taking \(\theta=\theta_J\) and using Eq.~\eqref{eq:sin_gap} proves
Eq.~\eqref{eq:robust_entangler_bound}. Moreover,
$L=4d=O(\Delta^{-1}\log(1/\eta))$.

Finally, we compute the total interaction time. Each primitive operator $R_{\phi}(\theta_J)$ uses free evolution for time $\tau$, so 
\begin{align}
     T_{\mathrm{ent}}=L\tau = O\!\left(\frac{1}{J_-}\log\frac{1}{\eta}\right),
\end{align}
where the last equality follows from $\Delta=\sin\theta_0 \geq\frac{2\theta_0}{\pi} =\frac{2J_-}{J_-+J_+}.$
\end{proof}

Applying Lemma~\ref{lem:robust_entangler} directly proves Theorem~\ref{thm:robust_swap}.

\begin{proof}
    Set $s=\operatorname{sgn}(J)$ and apply Lemma~\ref{lem:robust_entangler} with $\alpha=\pi/4$ and error $\eta/6$. Known single-qubit Clifford conjugations map $ZZ$ to $XX$ and $YY$, giving three $J$-independent sequences $\widetilde U_{PP}$ such that
\begin{equation}
    \left\|\widetilde U_{PP}
      -e^{-is\frac{\pi}{4}PP}\right\|\leq\frac{\eta}{6},
    \qquad P\in\{X,Y,Z\}.
    \label{eq:three_entanglers}
\end{equation}
The operators $XX$, $YY$, and $ZZ$ mutually commute and satisfy
\begin{equation}
    XX+YY+ZZ = 2\,\operatorname{SWAP}-I.
    \label{eq:swap_identity}
\end{equation}
Therefore
\begin{equation}
    \prod_{P\in\{X,Y,Z\}}e^{-is\frac{\pi}{4}PP}
    =e^{-is\frac{\pi}{4}(XX+YY+ZZ)}
    =e^{-is\pi/4}\operatorname{SWAP}.
    \label{eq:swap_decomposition}
\end{equation}
Write $U_{PP}:=e^{-is\pi PP/4}$ for the ideal rotations and define $\widetilde{\operatorname{SWAP}}(J):=\widetilde U_{XX}\widetilde U_{YY}\widetilde U_{ZZ}$. The triangle inequality and unitary invariance of the operator norm give
\begin{align}
    \left\| \widetilde{\operatorname{SWAP}}(J) - U_{XX}U_{YY}U_{ZZ} \right\|
    &\leq \sum_{P\in\{X,Y,Z\}} \left\| \widetilde U_{PP}-U_{PP}\right\|
    \nonumber\\
    &\leq \eta/2.
    \label{eq:swap_telescoping}
\end{align}
Let $\widetilde{\mathcal W}_J$ and $\mathcal W$ be the channels induced by $\widetilde{\operatorname{SWAP}}(J)$ and the ideal product, respectively. The inequality $\|\mathcal U-\mathcal V\|_\diamond\leq2\|U-V\|$ and Eq.~\eqref{eq:swap_telescoping} give $\|\widetilde{\mathcal W}_J-\mathcal W\|_\diamond\leq\eta$.

Finally, each of the three robust entangling rotations requires $O\!\left[ \frac{1}{J_-}\log\!\left(\frac{1}{\eta}\right) \right]$ interaction time. Therefore, the required interaction time for the SWAP gate is $ T_{\mathrm{SWAP}}(\eta) = O\!\left[ \frac{1}{J_-} \log\!\left(\frac{1}{\eta}\right) \right]$.
\end{proof}

\section{Proof for Lemma~\ref{lem:robust-marginal-distribution}}
\label{app:marginal-lemma}

The protocol contains mid-circuit measurements and resets, so we first specify the corresponding channel. Let $A$ be the qubit measured in a given round, let $B$ contain all unmeasured data qubits, and let $\rho_{AB}$ be their joint state immediately before the measurement. For an outcome $x$ in the chosen measurement basis, define
\begin{equation}
    p_x:=\operatorname{Tr}\!\left[\bra{x}_A\rho_{AB}\ket{x}_A\right]=\bra{x}\rho_A\ket{x},
    \qquad
    \overline\rho_B^{(x)}:=\frac{\bra{x}_A\rho_{AB}\ket{x}_A}{p_x}
    \quad (p_x>0).
    \label{eq:conditional-postmeasurement-state}
\end{equation}
Terms with $p_x=0$ may be assigned an arbitrary normalized state because they do not contribute. If the outcome is retained in a classical register $C$, the measurement channel is
\begin{equation}
    \mathcal M(\rho_{AB})
    =\sum_x p_x\ketbra{x}{x}_A\otimes\overline\rho_B^{(x)}\otimes\ketbra{x}{x}_C.
    \label{eq:measurement-channel-with-probability}
\end{equation}
Thus the post-measurement branch on $A$ is $p_x\ketbra{x}{x}_A=\bra{x}\rho_A\ket{x}\ketbra{x}{x}_A$, rather than the bare projector $\ketbra{x}{x}_A$. Resetting $A$ to $\ket0$ gives the completely positive trace-preserving channel
\begin{equation}
    \mathcal Q(\rho_{AB})
    :=(\mathcal R\circ\mathcal M)(\rho_{AB})
    =\ketbra{0}{0}_A\otimes\sum_x p_x\overline\rho_B^{(x)}\otimes\ketbra{x}{x}_C.
    \label{eq:measurement-reset-channel}
\end{equation}
Equivalently, one may use the unnormalized conditional operator $\rho_B^{(x)}:=\bra{x}_A\rho_{AB}\ket{x}_A=p_x\overline\rho_B^{(x)}$; the probability factor must not be omitted when $\overline\rho_B^{(x)}$ is normalized.

We next include the initialization error. Define the encoding unitary
\begin{equation}
    U_{A,k}:=e^{-iA_Zt_k}\prod_{j\in I_A}\mathrm{Had}_{r_j},
\end{equation}
where $\mathrm{Had}$ denotes the Hadamard gate. Also define the states immediately before alignment by
\begin{equation}
\begin{split}
    \rho_{\rm in}&:=U_{A,k}\widetilde\rho_0U_{A,k}^\dagger,\\
    \rho_{\rm in}^{\rm ideal}&:=U_{A,k}\ketbra0{0}^{\otimes N}U_{A,k}^\dagger
    =\ketbra{\Psi_A(t_k)}{\Psi_A(t_k)}.
\end{split}
\end{equation}
Unitary invariance and Eq.~\eqref{eq:initialization-trace-error} give
\begin{equation}
    \|\rho_{\rm in}-\rho_{\rm in}^{\rm ideal}\|_1\leq(N-1)^2\eta.
\end{equation}
Let $\mathcal C_m$ and $\widetilde{\mathcal C}_m$ be the ideal and implemented channels through measurement round $m$, including alignment and all measurement-and-reset maps $\mathcal Q_i$. There are at most $2m$ SWAP layers before this measurement: at most one alignment layer and $2(m-1)$ routing layers. For $c\in\{o,e\}$, let $\mathcal S_c$ and $\widetilde{\mathcal S}_c$ denote the channels of the layer unitaries $S_c$ and $\widetilde S_c$. Each layer contains at most $N/2$ disjoint SWAPs, so the triangle inequality gives
\begin{equation}
    \|\widetilde{\mathcal S}_c-\mathcal S_c\|_\diamond\leq\frac N2\eta.
    \label{eq:chain-layer-diamond-error}
\end{equation}
All intervening operations are channels and have diamond norm one. Applying the triangle inequality over the layers gives
\begin{equation}
    \|\widetilde{\mathcal C}_m-\mathcal C_m\|_\diamond\leq mN\eta.
    \label{eq:drain-channel-diamond-error}
\end{equation}
Combining the input-state and channel errors, and using contractivity under channels, yields
\begin{equation}
\begin{split}
    \|\widetilde{\mathcal C}_m(\rho_{\rm in})-\mathcal C_m(\rho_{\rm in}^{\rm ideal})\|_1
    &\leq\|\rho_{\rm in}-\rho_{\rm in}^{\rm ideal}\|_1
       +\|\widetilde{\mathcal C}_m-\mathcal C_m\|_\diamond\\
    &\leq(N-1)^2\eta+mN\eta.
\end{split}
\end{equation}
Tracing out every register except $C_m$ cannot increase this norm. The resulting classical states have trace-norm distance twice their total variation distance, so
\begin{equation}
    d_{\rm TV}(p_m,\widetilde p_m)
    \leq\frac12\bigl[(N-1)^2+mN\bigr]\eta
    \leq(N-1)^2\eta+2mN\eta.
\end{equation}
This proves Lemma~\ref{lem:robust-marginal-distribution}.

\paragraph{Independence between repetitions.}

Lemma~\ref{lem:robust-marginal-distribution} bounds the
marginal measurement error within a single experimental
shot. To estimate these marginals from repeated shots,
the sampling argument below uses independence between
repetitions. Under individual single-qubit control, independent
repetitions can be ensured by applying a fresh random
Pauli gate to every qubit before each initialization.
Choose each gate independently and uniformly from
$\{I,X,Y,Z\}$, and discard the random choices. The
resulting channel is
\begin{equation}
    \mathcal T(\rho)
    :=\frac{1}{4^N}
      \sum_{P\in\{I,X,Y,Z\}^{\otimes N}}P\rho P^\dagger
    =\frac{I}{2^N}.
\end{equation}
This identity holds for every input state, including
the register state conditioned on all preceding
measurement outcomes. Consequently, for fixed
Hamiltonian family, evolution time, and measurement
basis, every repetition begins from the same state
$I/2^N$ before robust initialization. With fresh
independent random choices in each repetition, the
resulting measurement outcomes are independent and
identically distributed across repetitions. The
initialization-error bound remains valid because it
holds for every input state. These additional Pauli
gates have negligible duration under the fast-control
convention and do not change the evolution-time cost. Under sublattice control, repetitions need not be independent, but this does not change the learning guarantees. The initialization and marginal-error bounds hold for every input state, including the state conditioned on all preceding measurement outcomes. Thus each conditional outcome probability remains within the same constant bias of its ideal Ramsey value. Applying Hoeffding's inequality for bounded martingale differences therefore gives the same asymptotic repetition count and total evolution-time bound.

\section{Proof for Theorem~\ref{thm:robust-parallel-learning}}
\label{app:robust-learning-proof}

We apply the robust phase-estimation guarantee of Sec.~\ref{subsec:rpe-main} to the $L=12N-9$ coefficient signals. For each coefficient $\lambda_a$, the ideal readout probabilities are $(1+\cos(2\lambda_at_k))/2$ and $(1+\sin(2\lambda_at_k))/2$, with $t_0=\pi/(4\Lambda)$ and $t_k=2^kt_0$. In particular, the initial scale is set by the bound on all coefficients, not merely by the upper bound $J_+$ on the backbone couplings.

Each coefficient is read in a round $m=O(N)$. Lemma~\ref{lem:robust-marginal-distribution} bounds the total variation distance between its ideal and implemented binary outcome distributions by $\beta_m:=[(N-1)^2+2mN]\eta$. For a binary distribution this also bounds the additive error in the probability of the $+1$ outcome. Choose $\eta=c/N^2$ with a sufficiently small fixed $c>0$ that this bias is at most $1/32$ in every round. Choose sampling accuracy $a=1/32$ for each probability. Equation~\eqref{eq:rpe-quadrature-error} then bounds the error of each estimated expectation by $2(a+\beta_m)\leq1/8$. The factor of two converts probability error to Pauli-expectation error.

Take $\gamma=\pi/6$ and let $K$ be the smallest nonnegative integer for which $t_K\geq\gamma/(2\varepsilon)$. By the branch-selection argument, successful sampling at every scale yields $|\widehat\lambda_a-\lambda_a|\leq\varepsilon$. Allocate failure probability to each coefficient and scale by
\begin{equation}
    \delta_k=\frac{\delta}{L}2^{k-K-1}.
    \label{eq:failure}
\end{equation}
With
\begin{equation}
    R_k=\left\lceil\frac{1}{2a^2}\log\frac{4}{\delta_k}\right\rceil
    =O\!\left(\log\frac{2L}{\delta}+K-k\right)
    \label{eq:repetition}
\end{equation}
independent repetitions per measurement basis and Hamiltonian family, Hoeffding's inequality controls the two quadratures together. A union bound over coefficients and scales gives total failure probability less than $\delta$.

It remains to count the Hamiltonian evolution time. By Theorem~\ref{thm:robust_swap}, each robust SWAP requires time $O(J_-^{-1}\log(1/\eta))=O(J_-^{-1}\log N)$. Disjoint SWAPs in a layer are executed simultaneously. Initialization and alignment and readout each require $O(N)$ layers. Thus their combined time per experiment is
\begin{equation}
    T_{\mathrm{init+read}}=O\!\left(\frac{N}{J_-}\log N\right).
\end{equation}
The $24$ Hamiltonian families and two measurement bases contribute a constant factor, giving
\begin{equation}
    T_{\mathrm{tot}}
    =O\!\left[\sum_{k=0}^KR_k(t_k+T_{\mathrm{init+read}})\right].
\end{equation}
Let $B=\log(L/\delta)$. The first term $\sum_kR_kt_k$ sums to
\begin{equation}
    \sum_{k=0}^KR_kt_k
    =O\!\left[t_K\sum_{j=0}^K2^{-j}
      \left(\log\frac{N}{\delta}+j\right)\right]
    =O\!\left(t_K\log\frac{N}{\delta}\right)=O(t_KB).
    \label{eq:rpe-sampling-time}
\end{equation}
Directly summing Eq.~\eqref{eq:repetition} gives $\sum_kR_k=O((K+1)B+K(K+1))$. Since $t_K=O(1/\varepsilon)$ and $K+1=O(\log(2\Lambda/\varepsilon))$, we obtain
\begin{equation}
\begin{split}
    T_{\mathrm{tot}}=O\!\left[
      \frac{1}{\varepsilon}\log\frac{N}{\delta}
      +\frac{N}{J_-}\log N
       \left(\log\frac{N}{\delta}+\log\frac{\Lambda}{\varepsilon}\right)
       \log\frac{\Lambda}{\varepsilon}\right].
\end{split}
\end{equation}
For $N\geq2$, replacing $L=12N-9$ by $N$ inside these logarithms changes only constants. This proves the stated accuracy, confidence, and time bounds. The physical scales are held fixed when stating the asymptotic dependence on $N$ and $\varepsilon$.

\section{Global control}\label{appendix:global_control}

In some experimental platforms, addressing individual dark qubits is difficult, whereas the same single-qubit gate can be applied simultaneously to a spatially periodic set of qubits. We briefly consider how the learning protocol can be adapted to this form of global control, focusing on a one-dimensional chain Hamiltonian in Eq.~\eqref{eq:parallel-general-H}.

Single-qubit control enters the protocol through Hamiltonian reshaping, local Clifford changes of basis, and the QSP construction of robust SWAP gates. To separate these issues, we assume here that the selected interaction component on every backbone edge is already of the form $g_i Z_iZ_{i+1}$, with $J_-\leq |g_i|\leq J_+$. The robust-SWAP construction then requires no additional Clifford transformation of the interaction axis.

We begin with Hamiltonian reshaping. Suppose that the control system can address the three sublattices of the chain separately. For $r\in\{0,1,2\}$, let
\begin{equation}
    \mathsf U_r:=\bigotimes_{\substack{j\in[N]\\j\equiv r\ ({\rm mod}\ 3)}}U_j
    \label{eq:period-three-global-control}
\end{equation}
denote the operation that applies the same single-qubit gate $U$ to every qubit in residue class $r$ and acts as the identity elsewhere. Sans-serif symbols distinguish these global operations from their single-qubit counterparts.

As a first step, twirl the sublattice $r=0$ by applying the four global Pauli operations $\mathsf I_0,\mathsf X_0,\mathsf Y_0,$ and $\mathsf Z_0$ with equal weight. Assuming for simplicity that $N$ is divisible by three, the resulting Hamiltonian is
\begin{equation}
\begin{split}
    H_{\mathrm{eff}}^{(0)}
    &:=\frac{1}{4}\sum_{P\in\{I,X,Y,Z\}}\mathsf P_0H\mathsf P_0^\dagger\\
    &=\sum_{q=0}^{N/3-1}\sum_{\alpha\in\{X,Y,Z\}}\left(h_{3q+1}^\alpha\sigma_{3q+1}^\alpha+h_{3q+2}^\alpha\sigma_{3q+2}^\alpha\right)\\
    &\phantom{={}}+\sum_{q=0}^{N/3-1}\sum_{\alpha,\beta\in\{X,Y,Z\}}J_{3q+1,3q+2}^{\alpha\beta}\sigma_{3q+1}^\alpha\sigma_{3q+2}^\beta.
    \label{eq:period-three-patch-reshaping}
\end{split}
\end{equation}
Every one- or two-body nearest-neighbor term intersects the sublattice $r=0$ on at most one qubit. The Pauli twirl removes any term acting nontrivially on that qubit and preserves every term supported entirely on the remaining sites. Consequently, Eq.~\eqref{eq:period-three-patch-reshaping} decomposes the chain into the disjoint two-qubit patches $(3q+1,3q+2)$. Choosing $r=1$ or $r=2$ translates the same patch construction along the chain. 

The first twirl retains all Pauli terms supported within each patch. To isolate one common Pauli component across the patches, we apply the same additional twirl to every patch. Fix a nonidentity operator $Q\in\{I,X,Y,Z\}^{\otimes 2}\setminus\{I\otimes I\}$ and let
\begin{equation}
    \mathcal C(Q):=\left\{P\otimes R\in\{I,X,Y,Z\}^{\otimes 2}:[P\otimes R,Q]=0\right\}
    \label{eq:patch-Pauli-centralizer-supp}
\end{equation}
be its Pauli centralizer, with irrelevant global phases omitted. Using the period-three controls on the two sites of every patch, define the further reshaped Hamiltonian
\begin{equation}
    H_Q:=\frac{1}{|\mathcal C(Q)|}\sum_{P\otimes R\in\mathcal C(Q)}\mathsf P_1\mathsf R_2H_{\mathrm{eff}}^{(0)}\mathsf R_2^\dagger\mathsf P_1^\dagger.
    \label{eq:global-centralizer-reshaping-supp}
\end{equation}
A Pauli operator survives this average only if it commutes with every element of $\mathcal C(Q)$. For a nonidentity two-qubit Pauli $Q$, the only such Pauli operators are $I\otimes I$ and $Q$. Since the Hamiltonian in Eq.~\eqref{eq:parallel-general-H} contains no identity component, the effective Hamiltonian therefore has the form
\begin{equation}
    H_Q=\sum_{q=0}^{N/3-1}\lambda_{q,Q}Q_{3q+1,3q+2},
    \label{eq:globally-isolated-Pauli-family-supp}
\end{equation}
where $Q_{3q+1,3q+2}$ denotes $Q$ acting on patch $(3q+1,3q+2)$ and $\lambda_{q,Q}$ is the corresponding coefficient $h_{3q+1}^\alpha$, $h_{3q+2}^\beta$, or $J_{3q+1,3q+2}^{\alpha\beta}$, according to the support of $Q$. Thus the same global pulse distribution isolates the same Pauli component on every disjoint patch while retaining its spatially varying coefficients.

Because the same operator $Q$ is selected on every patch, the required change of Pauli basis is also uniform. A pair of Clifford gates applied to the two corresponding sublattices maps the nonidentity factors of $Q$ to $Z$. After initialization through the bright qubit, sublattice gates prepare the target product state in the ideal circuit: one active qubit in each patch is prepared in $\ket{+}$ and, for a two-body term, the other is prepared in a $Z$ eigenstate. Evolution under $H_Q$ for time $t$ then encodes every $\lambda_{q,Q}$ simultaneously as a single-qubit phase record of the form $e^{-i\lambda_{q,Q}tZ}\ket{+}$.

It remains to transport these records to the bright qubit. Partition the nearest-neighbor edges into the three period-three matchings and define
\begin{equation}
    S_r^{(3)}:=\prod_{\substack{1\leq i<N\\i\equiv r\ ({\rm mod}\ 3)}}\operatorname{SWAP}_{i,i+1},\qquad r\in\{0,1,2\}.
    \label{eq:period-three-swap-layers}
\end{equation}
To implement one such layer, we reshape the Hamiltonian to the selected $ZZ$ backbone terms on the corresponding matching and apply the same QSP control sequence to all of its edges. The sequence depends only on $J_-$, $J_+$, and the target accuracy, so it remains valid when the coefficients $g_i$ vary from edge to edge. The robust-SWAP theorem therefore implements all SWAPs in $S_r^{(3)}$ simultaneously using only the available sublattice controls.

Initialization uses the same three sublattice SWAP layers. Apply a reset at site $1$ and then repeat $S_0^{(3)}$, $S_2^{(3)}$, $S_1^{(3)}$, and a reset $N-1$ times. The sweep contains each adjacent transposition once and forms a single cycle through all sites. To see the cycle property, start with isolated vertices and add these transpositions in their order of application. Every new chain edge joins two distinct connected components, and its transposition joins their permutation cycles. Thus the full sweep has one cycle. All original subsystems are consequently discarded after $N$ resets, even for an initially entangled state. Replacing the exact layers by the implemented layers gives $(N-1)^2$ approximate SWAPs and initialization error at most $(N-1)^2\eta$ in trace norm. The initialization depth is $3(N-1)$, so its native-Hamiltonian evolution time remains $O(NJ_-^{-1}\log N)$ for $\eta=O(N^{-2})$.

Consider first the patch arrangement in Eq.~\eqref{eq:period-three-patch-reshaping}. After a possible initial application of $S_1^{(3)}$, the phase records occupy sites $1,4,7,\ldots$, with the first record at the bright qubit $v_0=1$. Measure and reset this qubit, and then apply $S_0^{(3)}$, followed by $S_2^{(3)}$ and $S_1^{(3)}$. These three layers move the remaining records according to $4\to3\to2\to1$, $7\to6\to5\to4$, and so on, thereby restoring the same spatial pattern with the next record at the bright qubit. Repeating the measure--reset operation followed by this three-layer cycle reads all records in the selected family.

The two translated patch arrangements are handled in the same way, with at most three initial SWAP layers used to align their records with sites $1,4,7,\ldots$. Cycling over the three patch arrangements and the constant number of Pauli choices $Q$ covers every coefficient of the original Hamiltonian. Relative to the individually controlled protocol, global control therefore changes only the constant number of reshaping configurations and SWAP layers per readout round; the asymptotic total-time complexity remains unchanged.

For any fixed period $q\geq3$, the same construction uses pairs of adjacent sublattices as active patches and decouples the remaining sites. Cycling over the $q$ translated patch arrangements covers all coefficients. Define $S_r^{(q)}$ by selecting bonds whose left endpoint is congruent to $r$ modulo $q$. Each such bond set is a matching. Applying $S_0^{(q)}$, followed by $S_{q-1}^{(q)},\ldots,S_1^{(q)}$, moves each later record toward the bright qubit by $q$ sites when records occupy sites $1,1+q,1+2q,\ldots$. The same sweep contains every chain edge once and forms a single cycle, so it also initializes the register through repeated bright-qubit resets. For fixed $q$, the numbers of configurations and layers change only by constant factors.

\section{Tree routing with a single bright vertex}
\label{app:tree-routing}

This section gives the occupancy and error arguments used in Sec.~\ref{sec:general-graphs}.  The initial encoded state is a product of labelled phase-record factors and blank factors $\ket0$.  The labels are only a mathematical device for following the factors through exact SWAPs; the protocol does not measure whether a vertex is occupied by a record.

\subsection{Detailed proof of the rooted-tree drain}

We use the notation of Lemma~\ref{lem:tree-drain}. If the bright vertex initially holds a record, measure it and reset that vertex before beginning the schedule. Order the remaining sources $r_1,\ldots,r_\nu$ so that ancestors precede descendants. For each source, let $h_a:=\operatorname{lev}(r_a)$, and let $h_\star:=\max_a h_a$. The unique path from $r_a$ to the root is written as $r_a=x_{a,h_a},x_{a,h_a-1},\ldots,x_{a,0}=v_0$. Schedule the SWAP on $(x_{a,j},x_{a,j-1})$ in layer
\begin{equation}
 \tau(a,j)=h_\star+2a-j-1,
 \qquad 1\leq j\leq h_a,
 \label{eq:tree-drain-schedule-supplement}
\end{equation}
from Eq.~\eqref{eq:tree-drain-schedule}.  The first operation on record $a$ occurs in layer
\begin{equation}
 \tau(a,h_a)=h_\star-h_a+2a-1\geq1,
\end{equation}
so every operation has a positive layer index.

We first verify the matching constraint. Label a rooted edge by the level of its child. Two distinct edges incident on a common vertex are either sibling edges, whose child levels agree, or consecutive edges on a root path, whose child levels differ by one. Suppose events $(a,j)$ and $(a',j')$ occur in the same layer and their edges share a vertex. Equation~\eqref{eq:tree-drain-schedule-supplement} gives
\begin{equation}
 2(a-a')=j-j',
\end{equation}
with $|j-j'|\leq1$. Parity forces $j=j'$, after which $a=a'$. For fixed $(a,j)$, the root path of $r_a$ contains a unique edge whose child has level $j$. The events are therefore identical. Consequently the distinct edges scheduled in every layer are pairwise vertex disjoint.

We next prove that each scheduled SWAP carries the intended record toward the root. There are two points to establish. First, record $a$ remains at its source until layer $\tau(a,h_a)$. A different root path can touch $r_a$ only when its source is a descendant of $r_a$. If that source is $r_{a'}$, the ordering gives $a'>a$. The first event on that route incident on $r_a$ is the edge entering $r_a$, in layer
\begin{equation}
 \tau(a',h_a+1)
 =\tau(a,h_a)+2(a'-a)-1
 \geq \tau(a,h_a)+1.
 \label{eq:descendant-arrival-after-departure}
\end{equation}
Thus no other route touches the source before record $a$ begins to move.

Second, suppose inductively that record $a$ is at $x_{a,j}$ immediately before layer $\tau(a,j)$. Its next vertex is $x_{a,j-1}$. If this vertex still contained a record at its original source, that source would be an ancestor of $r_a$ and hence would have some index $a'<a$. Its scheduled departure layer satisfies
\begin{equation}
 \tau(a,j)-\tau(a',j-1)=2(a-a')-1\geq1,
 \label{eq:ancestor-departs-before-arrival}
\end{equation}
so it has already left. If instead $x_{a,j-1}$ contained another record currently in transit, that record must have entered in the preceding layer: every route, once started, uses consecutive layers. Its scheduled outgoing SWAP would therefore be incident on $x_{a,j-1}$ in layer $\tau(a,j)$, contradicting the matching property proved above. A record that has completed its route has been measured and the bright vertex reset. Hence $x_{a,j-1}$ is blank, and the scheduled SWAP moves record $a$ there.

Starting from Eq.~\eqref{eq:descendant-arrival-after-departure} and iterating this argument proves the claim for every $j=h_a,h_a-1,\ldots,1$. Record $a$ reaches the bright vertex in layer
\begin{equation}
 \tau(a,1)=h_\star+2a-2.
\end{equation}
Successive arrivals are separated by two layers. Under the fast-control convention of Sec.~\ref{subsec:access-model}, the arriving factor can be measured and the bright vertex reset before the next SWAP layer. Even if measurement and reset occupy one discrete routing layer, the intervening root-idle layer accommodates them. The final record arrives in layer $h_\star+2\nu-2$, proving Lemma~\ref{lem:tree-drain}.

\subsection{Initialization by a fixed tree sweep}
\label{app:tree-initialization}

We use the following result of D\'enes~\cite{Denes1959}.

\begin{theorem}[D\'enes]
\label{thm:tree-transposition-cycle}
Let $T=(V,E_T)$ be a tree with $|V|=N$. For each edge
$e=\{u,v\}$, let $\tau_e$ be the permutation of $V$ that
exchanges $u$ and $v$ and fixes all other vertices.
For any ordering $e_1,\ldots,e_{N-1}$ of the tree edges,
the product
\[
    \pi:=\tau_{e_{N-1}}\cdots\tau_{e_1}
\]
is a single cycle containing all $N$ vertices.
\end{theorem}

The sweep $D_T$ in Eq.~\eqref{eq:tree-initialization-sweep}
applies each tree-edge SWAP exactly once. SWAPs within
each matching commute, so their order within that layer
does not affect the sweep. Theorem~\ref{thm:tree-transposition-cycle}
therefore implies that the permutation $\pi$ induced by
$D_T$ is a single $N$-cycle.

Reset the bright qubit initially and after each of
$N-1$ sweeps. At reset round $k$, with $k=0,\ldots,N-1$,
the original subsystem discarded is the one initially
at $\pi^{-k}(v_0)$. Since
\[
    \{\pi^{-k}(v_0):k=0,\ldots,N-1\}=V,
\]
every original subsystem is discarded exactly once.
The resulting state is $\ketbra0{0}^{\otimes N}$.
This also holds for an initially entangled register,
because each reset discards the arriving subsystem
together with its correlations.

The implemented procedure replaces each SWAP by its
robust approximation. It uses $C_E(N-1)$ matching layers
and $(N-1)^2$ individual SWAPs. If each SWAP channel has
diamond-norm error at most $\eta$, the triangle inequality
and contractivity under resets give
\begin{equation}
    \|\widetilde\rho_0-\ketbra0{0}^{\otimes N}\|_1
    \leq (N-1)^2\eta.
\end{equation}
For bounded degree, $C_E=O(1)$, so initialization uses
$O(N)$ matching layers.

\subsection{Robust-SWAP error and time bounds}

Consider one selected Hamiltonian $A\in\mathfrak G$. It contains at most $N$ records, and its largest occupied level is at most $h_T\leq N-1$. Lemma~\ref{lem:tree-drain} therefore uses fewer than $3N$ matching layers. Let $\mathcal W_e$ and $\widetilde{\mathcal W}_e$ be the ideal and implemented SWAP channels on a tree edge, with
\begin{equation}
 \|\widetilde{\mathcal W}_e-\mathcal W_e\|_\diamond\leq\eta.
\end{equation}
For a matching $\mathcal M$, define the corresponding layer channels by
\begin{equation}
 \mathcal S_{\mathcal M}:=\prod_{e\in\mathcal M}\mathcal W_e,
 \qquad
 \widetilde{\mathcal S}_{\mathcal M}:=
 \prod_{e\in\mathcal M}\widetilde{\mathcal W}_e.
\end{equation}
The factors act on disjoint pairs.  The triangle inequality and stability of the diamond norm under composition with channels give
\begin{equation}
 \|\widetilde{\mathcal S}_{\mathcal M}
      -\mathcal S_{\mathcal M}\|_\diamond
 \leq |\mathcal M|\eta
 \leq \frac{N}{2}\eta.
 \label{eq:graph-layer-error}
\end{equation}

Apply the triangle inequality over the fewer than $3N$ layers preceding any readout. The initialization bound above contributes at most $(N-1)^2\eta$ in trace norm.  Measurement, reset, partial trace, and classical marginalization are channels and cannot increase trace distance.  Hence the total variation distance between the ideal and implemented marginal distribution of any record obeys the uniform bound
\begin{equation}
 d_{\rm TV}(p_a,\widetilde p_a)
 \leq 3N^2\eta.
 \label{eq:graph-marginal-error}
\end{equation}
The numerical constant is not optimized.  Choosing $\eta=c/N^2$ for a sufficiently small constant $c$ keeps this systematic bias below the constant threshold required by the robust phase estimator.

It remains to translate the exact routing depth into native-Hamiltonian evolution time. Every scheduled layer is a matching of edges in $E_T$. Reshaping isolates their designated Pauli components simultaneously, and individual Clifford gates map them to $ZZ$. Theorem~\ref{thm:robust_swap} implements every SWAP in that layer with error $\eta=O(N^{-2})$ using the same interaction time
\begin{equation}
 T_{\rm SWAP}=O\!\left(\frac{1}{J_-}\log N\right).
\end{equation}
The handshaking identity and the definition of maximum degree give
\begin{equation}
    2|E|=\sum_{v\in V}\deg(v)\leq N\Delta_G,
    \qquad |E|\leq\frac{N\Delta_G}{2}.
    \label{eq:edge-count-degree-bound}
\end{equation}
Using $C_V\leq\Delta_G+1$, $C_E\leq2\Delta_G-1$, $h_T\leq N-1$, and Eq.~\eqref{eq:edge-count-degree-bound} directly in Eq.~\eqref{eq:graph-routing-depth} yields the sharper form
\begin{equation}
    L_{\rm route}=O(N\Delta_G).
    \label{eq:graph-routing-depth-NDelta}
\end{equation}
Consequently, one complete sweep through all selected Hamiltonians at a fixed phase scale, measurement basis, and statistical repetition requires routing time
\begin{equation}
 T_{\rm route}
 =O\!\left(
 \frac{N\Delta_G}{J_-}\log N
 \right).
 \label{eq:graph-routing-native-time}
\end{equation}

Initialization is repeated for each selected Hamiltonian. Its contribution to a complete sweep is
\begin{equation}
    T_{\rm init}
    =O\!\left(\frac{(3C_V+9C_E)C_E N}{J_-}\log N\right)
    =O\!\left(\frac{N\Delta_G^2}{J_-}\log N\right).
\end{equation}
For bounded degree, both initialization and routing therefore take $O(NJ_-^{-1}\log N)$ time per complete sweep.

Equation~\eqref{eq:edge-count-degree-bound} also bounds the number of coefficients by
\begin{equation}
    3N+9|E|=O(N\Delta_G).
    \label{eq:graph-coefficient-count-NDelta}
\end{equation}
The same confidence allocation as in Eq.~\eqref{eq:repetition} therefore gives
\begin{equation}
 R_k=O\!\left(
 \log\frac{N\Delta_G}{\delta}+K-k
 \right).
\end{equation}
There are $3C_V+9C_E=O(\Delta_G)$ selected Hamiltonians and two measurement bases. Retaining the degree dependence before imposing the bounded-degree assumption, the total time is
\begin{equation}
 T_{\rm tot}
 =O\!\left[
 \sum_{k=0}^{K}R_k
 \left(\Delta_G t_k+\frac{N\Delta_G^2}{J_-}\log N\right)
 \right].
 \label{eq:graph-total-time-sum}
\end{equation}
Finally, $t_k=2^kt_0$, $t_K=\Theta(1/\varepsilon)$, and $K=O(\log(2\Lambda/\varepsilon))$. The sums evaluated in the proof of Theorem~\ref{thm:robust-parallel-learning} turn Eq.~\eqref{eq:graph-total-time-sum} into
\begin{equation}
\begin{split}
 T_{\rm tot}=O\!\Biggl[
 &\frac{\Delta_G}{\varepsilon}\log\!\left(\frac{N\Delta_G}{\delta}\right)\\
 &+\frac{N\Delta_G^2}{J_-}\log N\left(
 \log\!\left(\frac{N\Delta_G}{\delta}\right)
 +\log\!\left(\frac{2\Lambda}{\varepsilon}\right)
 \right)
 \log\!\left(\frac{2\Lambda}{\varepsilon}\right)
 \Biggr].
 \label{eq:graph-total-time-explicit-degree}
\end{split}
\end{equation}
The theorem assumes that $\Delta_G$ is bounded independently of $N$, so the additional factors of $\Delta_G$ in encoding and initialization are absorbed into the constants in big-$O$. This gives Eq.~\eqref{eq:robust-graph-learning-time}. For fixed physical and statistical parameters, the resulting scaling is $\widetilde O(N)$ on a bounded-degree graph. The same $O(N^{-2})$ gate-accuracy choice controls every graph readout marginal.

\end{document}